\documentclass[ reprint, amssymb, amsmath, aip,cha]{revtex4-1}
\usepackage[colorlinks=true,linkcolor=blue]{hyperref}%
\usepackage{graphicx}
\usepackage{amsmath,accents}
\usepackage{blkarray}
\usepackage{booktabs}
\usepackage{physics}

\newcommand{\vc}{\mathbf}

\newcommand{\etal}{{\em et~al.\/}}

\newcommand{\tildetext}{\raisebox{0.5ex}{\texttildelow}}

\usepackage{color}
\usepackage{soul}
\usepackage{amsmath,amssymb}%,wasysym}
\usepackage{graphicx}
\usepackage{empheq}
\usepackage{mathrsfs}
\usepackage{tikz}
\usetikzlibrary{shapes,arrows}
\usepackage{soul}
\setstcolor{red}

\begin{document}
\title{ ThunderBoltz: An Open-Source DSMC-based Boltzmann Solver for Plasma Transport, Chemical Kinetics, and 0D Plasma Modeling}
\author{Ryan Park}
\author{Brett S. Scheiner}
\thanks{Now at Lam Research Corporation}
\email{brett.s.scheiner@gmail.com}
\author{Mark C. Zammit}
\email{mczammit@lanl.gov}

\affiliation{Los Alamos National Laboratory, Los Alamos, NM, 87545}

\begin{abstract}
Plasma-neutral interactions, including reactive kinetics, are often either studied in 0D using ODE based descriptions, or in multi-dimensional fluid or particle based plasma codes. The latter case involves a complex assembly of procedures that are not always necessary to test effects of underlying physical models and mechanisms for particle-based descriptions. Here we present ThunderBoltz, a lightweight, publicly available 0D Direct Simulation Monte Carlo (DSMC) code designed to accommodate a generalized combination of species and arbitrary cross sections without the overhead of expensive field solves. It can efficiently produce high-quality electron, ion, and neutral velocity distributions in applied AC/DC $E$-field and static $B$-field scenarios. The code is built in the C++ standard library and includes a convenient Python interface that allows for input file generation (including compatibility with cross section data from the LXCat database), electron transport and reaction rate constant post-processing, input parameter constraint satisfaction, calculation scheduling, and diagnostic plotting. In this work we compare ThunderBoltz transport calculations against Bolsig+ calculations, benchmark test problems, and swarm experiment data, finding good agreement with all three in the appropriate field regimes. In addition to this, we present example use cases where the electron, ion, and background neutral particle species are self consistently evolved providing an ability to model the background kinetics, a feature that is absent in fixed background Monte Carlo and n-term Boltzmann solvers. The latter functionality allows for the possibility of particle-based chemical kinetics simulations of the plasma and neutral species and is a new alternative to ODE based approaches.
\end{abstract}
\maketitle

\section{Introduction}
Rate-equation-based chemical kinetic calculations have been widely used to help understand the variety of reactive and energy transfer processes encountered between electrons, ions, and neutral species in 0D models of plasma discharges\cite{refId0,Popov_2011,Popov_2016}. 
When employed for the description of plasma, chemical kinetics simulations track the energy transfer and species densities resulting from interactions between plasma electrons (described through coupling to a separate external Boltzmann equation solver code), neutral species, their electronic excited states, resulting dissociative products, and ions.  These descriptions represent the change of species by using a set of coupled ODEs in the form of species density conservation equations coupled by rate constants for each reaction process.
Often, a single species energy or temperature equation is used for all neutral species to track the consumption and deposition of energy due to the excitation or de-excitation of states, or endothermic and exothermic reactions, in heavy species collisions. Several open-source and commercial tools are available for such calculations including ZDplaskin\cite{ZDp}, CRANE\cite{keniley2019crane}, and COMSOL\cite{comsol}.  

These commonly used methods are limited in their ability to simultaneously treat electrons, ions, and neutrals with the energy or velocity distribution function (EDF or VDF, respectively) description that is afforded to electrons. Electron EDFs (EEDFs) are only able to react to changes in concentration, but cannot modify the EDF of ions or neutrals beyond their temperature, if it is tracked. These methods also fail to capture memory effects in the EDF. When coupling with an external Boltzmann solver it is assumed that the state of the EDF is parameterized by an $E/N$ (or mean energy) and species concentrations, assuming that the EDF has come into equilibrium on a time scale much shorter than the ODE integration time step. In this paper we present \emph{ThunderBoltz}, an Open-Source 0D Direct Simulation Monte-Carlo (DSMC) code that can overcome these difficulties by approximating solutions to the underlying coupled multi-species Boltzmann equation using a Monte-Carlo-based particle description for all charged and neutral species. 

\emph{ThunderBoltz} was conceived as a tool to study situations with strong plasma-gas coupling in the sense that the plasma modifies the gas composition and energy on a timescale comparable to the plasma dynamics. One such situation is the fast nanosecond scale collisional heating of neutral gas in electrostatic discharges by ion-neutral collisions and electron-neutral dissociation that leads to the formation of an outgoing shock wave ~100 ns after spark formation\cite{10.1063/1.3641413}. 
Due to the similarity with the shock formation in lightning discharges that results in thunder, the code was named \emph{ThunderBoltz}, \emph{Boltz} being short for Boltzmann. \emph{ThunderBoltz} is an Open-Source, portable, serial C++ code that is based on C++ standard libraries. The code makes use of a modification to the typical DSMC method where the velocity of charged particles is integrated in a background electric and magnetic field and updated each time step. This modification is akin to what is done in a Particle in Cell DSMC (PIC-DSMC) simulation\cite{6634908,VVarticle}, the difference being that the electric field in the present simulations is specified and not determined self-consistently by space charge distributions due to the 0D nature of the models it is intended to solve. ThunderBoltz is equipped with several collision models which can be evaluated using energy dependent cross section data provided by the user. An arbitrary number of particle species and collisions can be handled through the use of a user-friendly input deck, and the code at present can handle a static or frequency dependent electric field as well as a static magnetic field. The code is accompanied by a python API that increases the ease of use by rapidly generating input files, running simulations, and parsing cross section data from the LXCat database\cite{https://doi.org/10.1002/ppap.201600098}.  

Previously, a massively parallel 3D PIC-DSMC code was run for a single computational cell with fixed background N\textsubscript{2} particles in velocity space, fixed electric field, and no position update as a validation test\cite{VVarticle} to demonstrate agreement with Boltzmann solvers such as Bolsig+\cite{Hagelaar_2005}. A similar test has also been carried out in N\textsubscript{2}-O\textsubscript{2} gas in PIC Monte Carlo Collision simulations\cite{CHANRION20087222}. ThunderBoltz allows similar calculations to be run in a simple, portable, and easy to use 0D code without the overhead of machinery or learning curve needed to run or modify massively parallel 3D simulation codes. While the underlying method is common to PIC-DSMC, recognition of the value of using a DSMC simulation for 0D plasma modeling in place of a chemical kinetics type solver is new. As such, this paper focuses on the application of the method to solve benchmark and test cases, the comparison against the widely used two-term approximation numerical solver Bolsig+\cite{Hagelaar_2005} and swarm experiment data, and to the specific simulation parameter choices that are needed to obtain converged solutions. These results provide new insight into the accuracy of the DSMC collision method for modeling electron-neutral interactions in PIC-DSMC simulations, and the required time step and particle counts needed for converged results within a computational cell. In addition, we also focus on new applications to kinetics problems and 0D models, some of which have not been possible with the ODE approach. Two semi-realistic time dependent problems are considered to demonstrate the capabilities of the DSMC method that may be useful for particle-based kinetics modeling: the calculation of ion heating of neutral gas and its coupling to the ion mobility, and the time dependent calculation of the $\nu=0$--2 vibrational state density of N\textsubscript{2} in a sinusoidal time dependent electric field. In addition to this, we also demonstrate the inclusion of boundary interactions and the calculation of the breakdown voltage in nitrogen.

This paper is organized as follows: Section II presents details of the simulation methods and collision models implemented in the ThunderBoltz code. Section II also discusses convergence considerations for the time step and particle number. Section III demonstrates the use of ThunderBoltz on two benchmark plasma transport problems, and one neutral gas chemistry equilibrium test problem for which the solutions are known. Section IV gives example use cases including an example of a coupled ion, electron, and neutral transport problem where frequent interactions heats the neutral gas (Sec. IV A), a plasma chemical-kinetics type calculation (Sec. IV B), and the calculation of the Paschen curve (Sec. IV C). A discussion of results and future applications is given in Sec. V and a conclusion in Sec. VI.

\section{Simulation Methods}
\begin{figure}
\begin{tikzpicture}[node distance = 3cm, auto]
\tikzstyle{block} = [rectangle, draw, fill=white!20, 
text width=5.5em, text centered, rounded corners, minimum height=4em]
\tikzstyle{line} = [draw, -latex']

\node [block] (1) {Every $N_{\rm sample}$ steps sample $\big[c\sigma(c)\big]_{\rm max}$};
\node [block, below right of=1] (2) {Evaluate $N_{\rm pairs}$ for the set of $p$ collisions};
\node [block, below left of=1] (3) {Create products, remove reactants};
\node [block, above left of=3] (4) {Accelerate in $\vc{E}$ and $\vc{B}$};

\path [line] (1) -- (2);
\path [line] (2) -- (3);
\path [line] (3) -- (4);
\path [line] (4) -- (1);
\end{tikzpicture}
\caption{Simulation flow}
\label{fig:loop}
\end{figure}
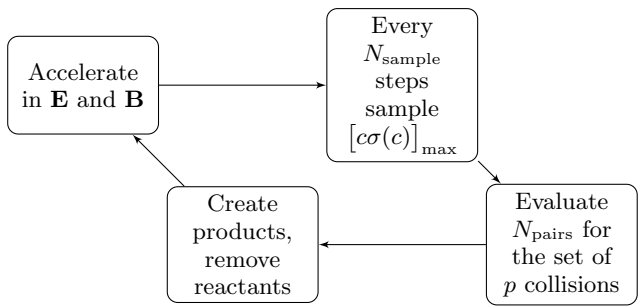

This section describes the numerical methods employed by ThunderBoltz and the various use cases and convergence considerations.
%\subsection{No-time-counter with Boris integration}
As shown in Fig.~\ref{fig:loop}, ThunderBoltz makes use of a slight modification of the DSMC no-time-counter (NTC) method\cite{bird1994molecular}, with an additional acceleration step for charged particle motion in an $E$ or $B$ field, and periodic re-sampling of $\left[c\sigma_p(c)\right]_{\rm max}$ described below. 
The main simulation loop starts with the acceleration step where particle velocities are updated using either a forward difference approximation in the absence of an external magnetic field, or the Boris velocity integration method \cite{boris1970acceleration} if a magnetic field is present. 
The cumulative particle displacement is tracked in the 0D simulation in a similar sense to what is done for Monte Carlo Boltzmann simulations\cite{White_1997,DIAS2023108554} for the purpose of calculating bulk transport coefficients,  but the displacement is not considered in determining the likelihood of particle interaction.

The DSMC NTC method approximates solutions to the Boltzmann equation by sampling collisions based on energy dependent  collision cross section data between particles present in the simulation. In 3D DSMC simulations particle collision pairs are selected on a cell by cell basis, however, in 0D simulations all particle pairs within the simulation are considered for evaluation. 
For a collision process, $p$,
involving species of type $A$ and $B$,
the number of interaction pairs specified by the NTC method is 
\begin{equation}
N_{\rm pairs}^{p}=\left(1-\frac{\delta_{AB}}{2}\right) \frac{N_A N_B W [c\sigma_{p}(c)]_{\rm max}\Delta t}{L^3}\label{NTC}.
\end{equation}
Here  
$N_s$ is the number of particles of species $s$,
$\Delta t$ is the time step, 
$W$ is the particle weight,  
$L$ is the length of a cell, 
$\delta_{AB}$ is the Kronecker delta,
$\sigma_{p}$ is the collision cross section, and
$c_{ij}$
is the relative velocity of a selected particle pair,
where $i \in [1,N_A]$ runs over the particles of type $A$, and
$j \in [1,N_B]$ runs over the particles of type $B$.
The quantity $[c\sigma_{p}(c)]_{\rm max}$ is the maximum product of the cross
section and relative velocity 
for interaction pair samples
within the ensemble. 
%for a selected interaction pair. 
Due to the arbitrary volume $L^3$ of a 0D simulation, physical weighting of particles
$W=1$ is used. 
For each process, $p$, $N_{\rm pairs}^{p}$ pairs of
particles of type $A$ and $B$ are randomly selected and the probability of a collision for each particle pair
$A_i,B_j$ is computed as 
\begin{equation}
P_{ij}^{p}=\frac{ c_{ij}\sigma_{p}(c_{ij})}{\left[c\sigma_{p}(c)\right]_{\rm max}}.
\end{equation}  
A collision is accepted if $R<P^{p}_{ij}$,
where $R$ is a random number between 0 and 1, 
which is taken from a uniform distribution. 

Typically,  $[c\sigma_{p}(c)]_{\rm max}$ is estimated by sampling values from the total ensemble of possible interaction pairs for each reaction. In principal the exact value of $[c\sigma_{p}(c)]_{\rm max}$ does not change the collision frequency, but it does affect the number of particles evaluated for collision and may lead to computational inefficiency depending on the value used. In many applications of the NTC method this value is not sampled every time step because it has a tendency to change very little. Due to the change in distribution function of charged particles reacting to acceleration in a strong electric field, the option for resampling $[c\sigma_{p}(c)]_{\rm max}$ is included in the simulation loop.  

\begin{figure}
    \centering
    \includegraphics[width=.48\textwidth]{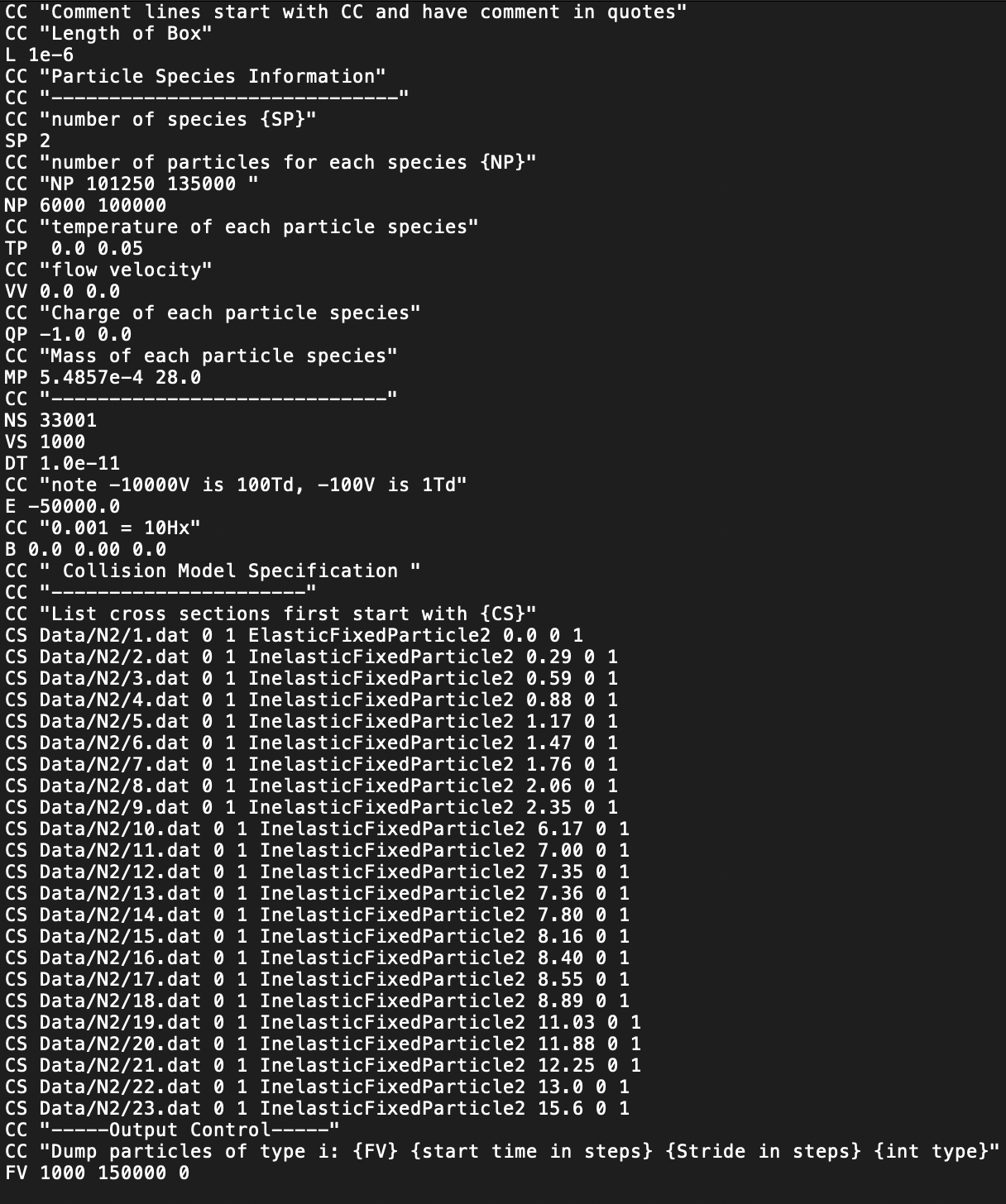}
    \caption{An example input deck for calculating rate constants and transport for electrons colliding with nitrogen gas. Two species and 23 cross sections are included in the example.}
    \label{fig:input}
\end{figure}

ThunderBoltz simulations are setup using a text based input file, shown in Fig.~\ref{fig:input} that includes the specification of an arbitrary number of particle species with their mass, charge, initial temperature, and flow velocity. Each species is identified by an assigned integer, in order of specification, starting with the electron species at 0. The input is structured so that a two letter key starts the line that specifies the input parameter(s). The input includes the specification of an accelerating constant or time varying electric field in the z direction, a DC magnetic field and arbitrary direction, time step, $[c\sigma_{p}(c)]_{\rm max}$ time step sample interval, number of time steps, and collisions. The input deck allows for a variety collisions and collision models. The collision specification line starts with the two character identifier ``CS" followed by a path to an energy dependent cross section data file, two integers specifying each of the colliding particle species, the collision model name, a threshold energy for the interaction in eV, and two integers specifying the reaction products. The variety of options for collision specification is the core feature that provides flexibility for use in multiple applications. Various use cases and collision models are described in the following subsections.     

\subsection{0D Simulation Use Cases\label{sec:use}}
This subsection outlines various use cases of 0D kinetic simulations that are presented throughout this paper. When the background population is fixed, some of these are analogous to the typical use of Monte-Carlo Boltzmann solvers, while others employ different assumptions relating to how the charged and background species interact. Three categories of simulations are outlined below. This list is non-exhaustive and other combinations of various aspects of these use cases are possible with the included collision models.

\subsubsection{Charged Particle Transport}

The first category of use is the calculation of charged particle transport. This closely parallels the use of most Monte-Carlo Boltzmann solvers\cite{Tejero-del-Caz_2019,RABIE2016268,DIAS2023108554}, the primary difference being that the background population is represented by particles. To emulate the infinite uniform background in MC simulations in this scenario, charged particles interact with the neutral background particles, but the post collision velocities and species of the background particles are not updated. In these cases, excitation processes of the background gas are modeled as inelastic loss processes so that charged particle energy loss occurs, but the products are not tracked. In addition to inelastic loss, the code also provides options for turning electron generation on or off during ionization events. In the former case electrons are produced in electron impact ionization events, while in the latter case, ionization is treated as an inelastic loss process with no additional particles generated. Collision models for these application are listed in Sec.~\ref{sec:collmodel}.

The calculation of bulk and flux transport quantities is identical to that which has been presented previously for MC simulations\cite{DIAS2023108554}. Here, the definitions of rate constants, mean energy, bulk and flux flow moments, mobility, and Townsend ionization coefficients are reviewed since these are used in the benchmark tests and swarm experiment data comparisons presented in Sec. III.

Standard output at each output step includes species densities, temperatures, flow velocity, mean energy, cumulative particle displacement (for electrons), and the cumulative reaction count $C_p$ for each interaction specified. These quantities allow the calculation of all transport quantities described in this section.
For example, 
reaction rates constants, $k_p$,
can be computed in post-processing using the relation
\begin{equation}\label{eq:rates}
k_p=\frac{1}{L^3 n_A n_B}\frac{dC_p}{dt},
\end{equation}
where $n_s$ is the $s$ species particle number density.
 Likewise, energy loss rates for each reaction can also be calculated by multiplying by the threshold energy. The flow velocity can be defined either as a flux quantity by taking the appropriate moment of the species, $s$, velocity distribution function, $f_s(\vc{v})$:
\begin{equation}\label{eq:vflux}
    \vc{V}_s^{\textrm{Flux}}=\langle \vc{v} \rangle \equiv \frac{1}{N_s}\int \vc{v} f_s(\vc{v}) d^3v,
\end{equation}
or as the velocity defined by the time derivative of the mean displacement of the charged particle swarm: 
\begin{equation}\label{eq:bflux}
    \vc{V}_s^{\textrm{Bulk}}=\frac{d\langle \vc{r} \rangle}{dt}.
\end{equation}
The flux flow velocity moment is computed with the following sum of simulated particle quantities:
\begin{equation}\label{eq:vdirectional}
V_{s,u}=\frac{1}{N_s}\sum_{i\in s}^{N_s} v_i\cdot\hat{u},
\end{equation}
where %$N_e$ is the number of electrons, 
$\hat{u}$ is the unit vector in the $u$ direction, and $v_i$ is the velocity of particle $i$ in the laboratory frame. 
%\hl{MZ: This needs rephrasing: This variable is not limited to bulk calculations} 
For the calculation of bulk transport coefficients the swarm displacement is needed. 
For this purpose, each particle has a variable $\vc{r}_i$ that tracks its cumulative displacement.
The mean displacement vector averaged over all particles of a given species $s$ is also provided in the output
\begin{equation}
    \langle\vc{r}_s\rangle=\frac{1}{N_s}\sum_{i\in s}^{N_s} \vc{r}_i,
\end{equation}
allowing computation of Eq.~\eqref{eq:bflux}.
The two flow velocity definitions are useful for calculating bulk and flux transport coefficients\cite{PT-Casey_2021,DIAS2023108554}. Another quantity of interest included in the standard species output is the mean energy,
\begin{equation}\label{eq:meanE}
\langle\mathcal{E}_s\rangle=\frac{m_s}{2N_s}\sum_{i\in s}^{N_s}  v_i\cdot v_i,
\end{equation}
which is often a preferable dependent variable for the tabulation of transport quantities\cite{PhysRevE.80.036405}.

In Sec.~\ref{sec:HeSwarm}, the bulk and flux calculations for the mobility and Townsend ionization coefficient are calculated for electrons in Helium gas using the phenomenological relation 
\begin{equation}\label{eq:mobility}
    \vc{V}_s=\mu_s \vc{E},
\end{equation}
where $\mu_s$ is the species mobility, and
\begin{equation}\label{eq:alpha}
    \frac{\alpha}{N_{\rm gas}}=\frac{k_{\rm ionization}}{\vc{V}_s\cdot\hat{z}},
\end{equation} along with the appropriate bulk or flux definition of $\vc{V}_s$, and are compared with swarm data and Bolsig+ where available. 

Here we have neglected the calculation of the diffusion tensor, however, this quantity (including parallel, perpendicular, and Hall terms) can be trivially included by adding output for $\langle x^2\rangle$, $\langle y^2\rangle$, $\langle z^2\rangle$, $\langle xy\rangle$, $\langle xz\rangle$, and $\langle yz\rangle$. This and other updates will be included in future versions of the code.

\subsubsection{0D Reactive Kinetics}
The reactive kinetics simulations do not utilize a background species. All particles interact and react with each other as specified by the input collision list. It is possible to calculate charged particle transport in this scenario, however, there is no guarantee that the neutral species are at a steady-state density and transport properties may change as species are populated or depopulated through various reaction pathways. It is also possible for the neutral species to gain energy from the field accelerated particles either directly through elastic collisions, via exothermic reactions (reactions with negative threshold energies), or from the excess energy after a dissociation process. 
%\hl{MZ: Dissociation reactions can produce 'hot' neutrals. Not sure how much you want to go into this}. 
Examples of this type of simulation are given in Sec.~\ref{sec:gasHeating} and Sec.~\ref{sec:kinetics}. 

Often, it is desirable to include boundary interactions in 0D models. While none have been included in the current version of the code, their implementation is simple. See Sec.~\ref{sec:PaschenCurve} and Sec.~\ref{sec:disp} for relevant discussion.

\subsubsection{Fixed Ionization Fraction}
In some situations it is desirable to allow some aspects of plasma-neutral interactions to be modeled, while neglecting others. One such circumstance involves questions regarding the rate of heating of neutral gas due to ion neutral and electron neutral collisions and how this rate changes with ionization fraction. In this case, the energy exchange between particles is allowed but the electron-neutral reaction products are not tracked and are treated as an inelastic loss process. An example of this is given in Sec.~\ref{sec:gasHeating}.  

\subsection{Collision Models\label{sec:collmodel}}
This section outlines the various collision models that have been implemented in the code. The motivation for many of these was given in the use cases in Sec.~\ref{sec:use} and will be elaborated on in this subsection.  

\subsubsection{Isotropic Elastic Collisions} 
In this collision model the energy and momentum exchange between particles is calculated assuming a uniformly random isotropic post-collision scattering angle. 
%\hl{MZ: Why?:} \hlcy{BSS: Because the title of this subsubsection is "Isotropic Elastic Collisions"?}
This model is intended to be used with elastic momentum cross section data, 
where in  many cross section databases  momentum transfer cross sections are included instead of elastic cross sections. 
%\hl{MZ: This is not what is done in LXCat, did you have something else in mind?:} \hlcy{RP: Maybe we clarify that this is how the MTCS is defined, rather than this is how it was obtained, since you could obtain it from many different sources.}\st{Momentum scattering cross sections have taken an angular average over the anisotropic differential scattering cross section to provide an effective isotropic cross section for momentum transfer.} 
These can  be used in conjunction with the isotropic  and elastic collision model to produce transport data that captures the momentum transfer  statistics
%\hlcy{RP: The expected momentum transfer, $\langle \Delta \Vec{p} \rangle_{ij}$, of any two particles $i$,$j$ is $\frac{c_{ij}\sigma_{\rm MTCS}\Delta t}{L^3}\Vec{p}_0$. So, the average mom. transfer is correct, but this compensation results in an artificial  collision rate.} 
of the anisotropic scattering when models for the anisotropic scattering angle distribution are unavailable.
  
\subsubsection{Anisotropic Elastic Collisions}

%ThunderBoltz allows for the implementation of arbitrary inverted angular distribution functions, which relate the post-collision scattering angle to a uniform random number between 0 and 1. At present, ThunderBoltz contains the Park inverted angular distribution function form \cite{Park_2022}, which was recently utilized to reproduce all-order electron elastic scattering differential cross section of H, He, and H$_2$. 
%\hl{MZ: Ryan, this sentence is not clear and may need a bit more details to set it up. For the second sentence of the paragraph we should continue the discussion of how the code handles the anisotropic models and does rotations and then go into details of the Park model and its limitations OR we have a new section that can outline possibilities of including new models} 
%The post-collision relative velocity vector $\bf{g^\prime}$ is initially shifted, to maintain consistency with angular distribution referenced to the initial relative velocity vector $\bf{g}$, $\bf{g^\prime}$ is rotated into the lab frame so that the post-collision scattering angles are correctly referenced to $\bf{g}$. See the detailed discussion in Ref.~\onlinecite{donko2021} for details of the procedure. Since the angular scattering is included in the model, it is only intended for use with elastic scattering cross sections.

ThunderBoltz includes the framework needed for the implementation of anisotropic scattering models. Generally, independent of the particular anisotropic model, the post collision scattering angle is determined in the center of mass frame. The relative velocity vector is rotated in the center of mass frame throughout the collision and needs to be transformed back to the original laboratory (simulation) frame to determine the post collision velocities. The mechanism for performing this transformation is available in the code for the implementation of anisotropic models. See the discussion in Ref.~\onlinecite{donko2021} for details of the procedure.  

At present, ThunderBoltz includes the anisotropic elastic scattering model from Ref.~\onlinecite{Park_2022}. The model provides an invertible angular distribution function that allows the post collision scattering angle of an electron to be determined by its relative collision energy and a uniform random number between 0 and 1 such that the scattering angle distribution of electrons approximates that given by the elastic differential scattering cross section when collisions are evaluated with an appropriate elastic integrated scattering cross section. The model currently has known fit parameters for the inverted angular distribution function and the elastic integrated scattering cross section of H, He, H$_2$, and their isotopes.

\subsubsection{Non-Reactive Collisions} 
\label{Sect:Non-Reactive_Collision}
This collision option can be combined with inelastic collisions and neglects the products produced in reactions, i.e the reactants are the same as the products.  
The appropriate energy transfer or threshold energy cost of the reaction is calculated and subtracted from the reactants post collision velocities. 
This collision type is useful for emulating the behavior of Boltzmann solvers by maintaining an unchanging background, or when the tracking of excited states of some subset of species in reactive kinetics simulations is unimportant. As an example, if the final density of   N$_2$($A^3\Sigma_u^+$) is unimportant, 
the reaction e~+~N$_2$~$\to$~e~+~N$_2$($A^3\Sigma_u^+$) is modeled as e~+~N$_2$~$\to$~e~+~N$_2$ with the excitation energy of the $A^3\Sigma_u^+$ state subtracted from the relative collision energy, but no particles are removed and no new particle products are generated.

\subsubsection{Fixed Heavy Particle Collisions} 
\label{Sect:Fixed_Heavy_Particle_Collision}
To calculate transport of a charged species in a fixed background gas it is necessary to maintain the statistical properties of the background (e.g. temperature, energy, flow and other moments of the VDF). If energy exchange in elastic or inelastic collisions is allowed the background will gain energy from the field accelerated particles. For this reason, it is desirable to have a collision model that fixes the background particles. In this case, the electron or ion energy loss is calculated as in an elastic collision, but the neutral heavy particle energy and velocity is not updated, maintaining the statistical properties of the background gas distribution.

\subsubsection{Isotropic Inelastic Collisions} The energy and momentum exchange between particles is calculated using a uniformly random isotropic post collision scattering angle and the threshold energy for the inelastic collision process is subtracted in the center of mass frame. 
This model can be combined with the Fixed Heavy Particle and Non-Reactive Collision options as described in Secs.~\ref{Sect:Non-Reactive_Collision}
and
\ref{Sect:Fixed_Heavy_Particle_Collision}, 
or
the species type of the reactants is switched to that of the products given in the collision specification line. Other than ionization, currently all inelastic collisions use this scattering model.

\subsubsection{Electron Impact Ionization}
For ionizing collisions, a model is needed to determine how to partition energy in excess of the ionization energy between the products. The residual ion maintains the velocity and direction of the target neutral/ion, while the scattering angle of the product electrons is assumed to be isotropic. The excess energy available for sharing between the electrons is calculated by assuming the post-collision residual ion maintains the same velocity
as the pre-collision target neutral/ion. The energy balance in the rest frame of the target is $\epsilon_{\rm s} = \epsilon - \epsilon_{\rm ion}-\epsilon_{\rm ej}$,
where $\epsilon$, $\epsilon_{\rm s}$, and $\epsilon_{\rm ej}$ are the
incident, scattered, and ejected electron energies, and $\epsilon_{\rm ion}$ is the ionization energy of the bound electron. Generally, the ejected electron can take on a distribution of values which can be represented by an electron energy sharing distribution. Three models for this distribution are implemented.
These are the \emph{one takes all} model: where one electron is ejected with 0 eV and the other with $\epsilon - \epsilon_{\rm ion}$\cite{Hagelaar_2005}, the \emph{equal energy sharing} model\cite{Hagelaar_2005}: where each electron is ejected with energy $(\epsilon - \epsilon_{\rm ion})/2$, and the \emph{uniform energy sharing} model\cite{CHUNG20053}: where the energy of one electron has a uniform distribution in the range $[0,(\epsilon - \epsilon_{\rm ion})/2]$ and the other is determined from conservation of energy.

%\subsubsection{Anisotropic Elastic Modeling
%One advantage of the direct simulation methods is the ability to easily 
%implement anisotropic scattering models without a truncated angular basis
%expansion. The current version of the code allows for the specification
%of an arbitrary inverted angular distribution function, $\cos\chi(\epsilon, R)$,
%which relates the post-collision scattering angle to the relative energy
%of the colliding species and a random number between 0 and 1 to be re-sampled
%for each collision. The code implements an inverted angular distribution
%from Park~\etal~\cite{Park_2022} which has been optimized for He, H, and H$_2$.

%In the case of isotropic collisions, the orientation of the initial relative velocity
%vector is irrelevant (since isotropic distributions are invariant under rotation). However,
%this is not the case with anisotropic collisions, and one must be careful to rotate the
%post-collision velocities back into the common simulation frame. We follow the procedure
%specified by Donk\'o~\etal~\cite{donko2021} to do this.

\subsection{Convergence Considerations}
\label{conv_cons}

Certain 0D kinetic simulation parameters (time step, total particle count, and cell size) should not
affect the resulting calculation of reduced steady-state transport coefficients.
%Traditionally, these parameters each require independent convergence testing at various reduced field
%strengths for each background system of interest. Such investigations are further complicated by
%computational resource limitations and nonlinear relationships between the input conditions.
Here, we introduce some considerations for the DSMC-NTC method that will aid the process of
obtaining converged results.

(1) Enough collisions should occur per time step that particles do not unphysically run away.

(2) The simulation time step should be short enough to allow for sufficient resolution of the input cross sections.

% (3) Collision should not be so frequent that pairs of particles often get sampled multiple times during the
% same time step, or that the electron VDF relaxes considerably between time steps. \hl{Get rid of this? Should we
% have this bullet point instead preface the Stosszahlansatz subsection?}

Details for establishing each of these criteria are investigated in the following sections, then the consequent relations
to time step, cell size, and particle number specifications are presented.

\subsubsection{Lower Bound of $N_{\rm pairs}$}

It has been found that enough collision pairs must be considered in order to get
the correct statistical results. If collision probabilities are too small for important
collision processes, then the cross sections will not be properly sampled. That is,
each process should satisfy,
\begin{equation}
    N_{\rm pairs}^{p} \geq N_{\rm min},
    \label{Pcrit}
\end{equation}
where $N_{\rm min}$ is a new free
parameter that must be increased until convergence is seen in the transport parameters
of interest. For the tests in this paper, a value of $\tildetext$100 was sufficient to
obtain convergence of mean energy, reduced electron mobility, and reduced ionization
rate.

%\subsubsection{Upper bounds on N_{\rm pairs}}

%We can impose an upper bound on the number of collisions each species should be considered for
%during each time step. In particular, the total pairs of collisions should not be roughly greater
%than the number of particles of that species

%\begin{equation}
%    \sum_{B,i} N_{{\rm pairs}, A, B, i} < C N_A.
%    \label{Ccrit}
%\end{equation}
%Here, we have some scalar $C$ that we can be decreased if collision pairs are being
%selected multiple times. \hl{the probability for pairs being considered twice is proportional to $\left(\frac{N_{pairs}}{N_AN_B}\right)^2$}

\subsubsection{Cross Section Energy Resolution}
\begin{figure}[hbt!]
\centering
% \includegraphics[width=.5\textwidth]{TimeStep.pdf}
% \caption{\label{Fig:CX} A figure displaying cross sections from the TRINITI Database\cite{TRINITI} obtained from LXCat\cite{https://doi.org/10.1002/ppap.201600098}. The figure is annotated to demonstrate the energy gain of an electron initially at zero velocity in a field of E=100V/m during time steps of 10, 100, and 1000 ns.   }
\includegraphics[width=.5\textwidth]{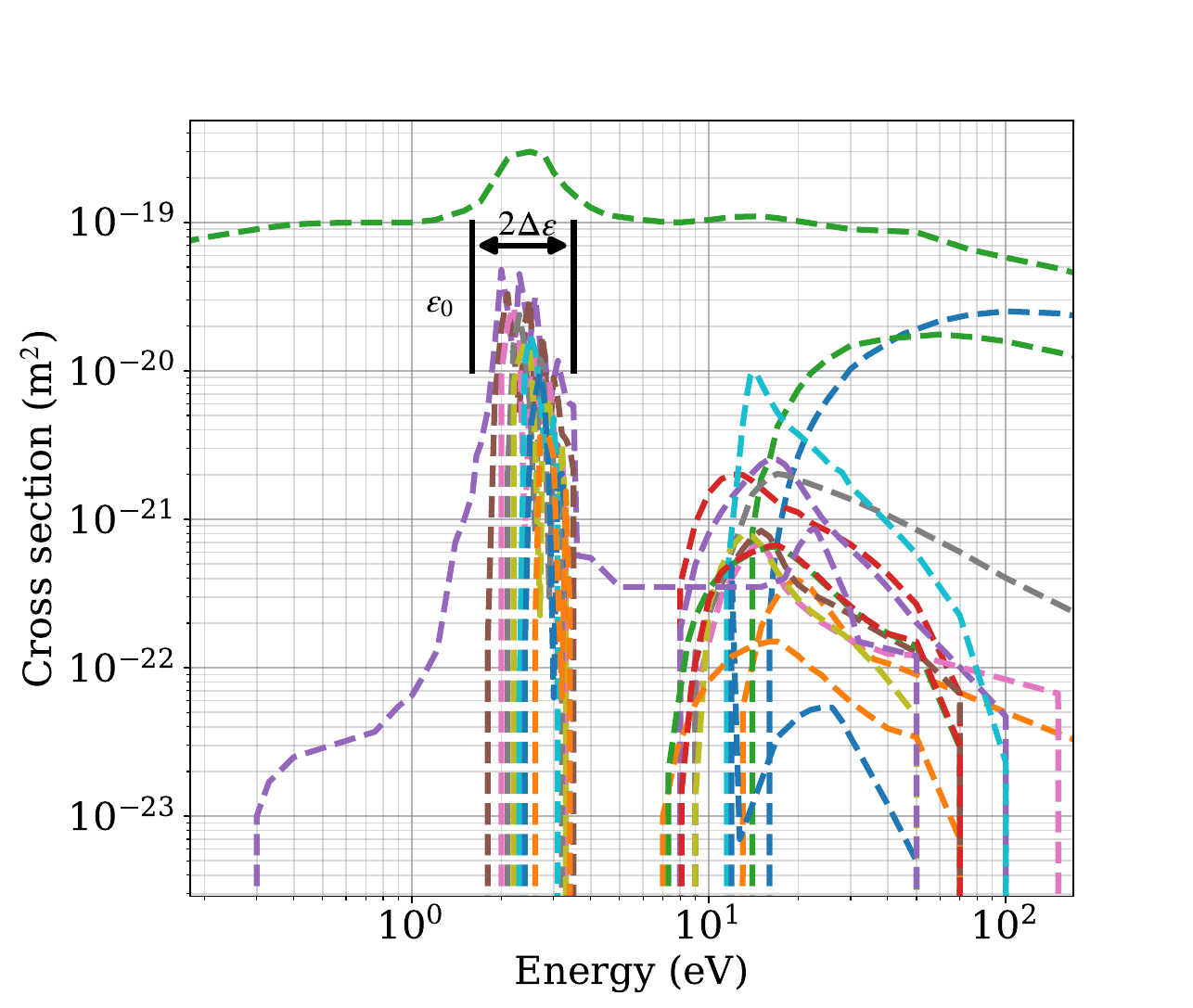}
\caption{\label{Fig:CX} A figure displaying N$_2$ cross sections from the TRINITI Database\cite{TRINITI}
obtained from LXCat\cite{https://doi.org/10.1002/ppap.201600098}. The annotations demonstrate how to obtain
minimum energy gain criteria by observing prominent cross section peaks.
}
\end{figure}

The time step for the evaluation of the simulation loop is chosen so that collisions are sufficiently sampled between the electric field particle accelerations. 
This time step requirement is driven by a need to place a limitation on the energy gain
by a single particle during one step so that all features of the cross section are sampled
as a particle accelerates through a range of energies. This requirement has been extensively
discussed by Moore and coauthors in the context of PIC-DSMC simulations\cite{2015APS..GECQR4004M}.
The requirement is not associated with a hard and fast rule, but needs to be evaluated for
a given set of cross sections and a given accelerating electric field. To do this, one could
identify some worst-case scenarios. For example, a non-colliding electron travelling in the
field direction will experience the following change in energy during time step $\Delta t$,
\begin{equation}
    \Delta \epsilon(\Delta t) = q_eE\Delta t\sqrt{\frac{2\epsilon_0}{m_{\rm e}}}
        + \frac{1}{2}\frac{q_e^2E^2}{m_{\rm e}}\Delta t^2, \\
    \label{DE}
\end{equation}
where $q_e$ is the unsigned electron charge, $m_e$ is the electron mass, $E$ is the electric field magnitude,
and $\epsilon_0$ is the electron initial energy. 
One can intuitively assign a value of 
$\epsilon_0$ as the 
approximate mean energy of the EDF.
Then, by observing a cross section profile such as that in Fig.~\ref{Fig:CX}, 
%one can intuitively assign values for $\Delta \epsilon$ and $\epsilon_0$ that are suitable for sampling prominent cross section features. 
one can intuitively assign a value of $\Delta \epsilon$  that is suitable for sampling prominent cross section features. 
Assuming $\Delta t^2$ is negligible,
this imposes a condition on $\Delta t$,
\begin{equation}
    \Delta t = \frac{\Delta \epsilon}{q_eE}\sqrt{\frac{m_{\rm e}}{2\epsilon_0}}.
    \label{DTcrit}
\end{equation}

Often, we would like to specify the working conditions within units of the reduced electric field,
and so the electric field itself is a function of the cell size and the input reduced field,
\begin{equation}
    E[{\rm V/m}] = 10^{-21}\frac{(E/n_{\rm gas})[{\rm Td}]N_{\rm gas}}{L^3},
    \label{Efield}
\end{equation}
where $N_{\rm gas}$ and $n_{\rm gas}$ are the gas particle number and gas particle number density, respectively.

By combining the requirements from Eqs.~\eqref{Pcrit}, \eqref{DTcrit}, and \eqref{Efield}, we reach a condition on the particle
counts of each reaction pair such that at least $N_{\rm min}$ pairs are sampled during the collision routine.
This allows one to write 
a condition for the number of particles  
such that

%\hl{MZ: Ryan, please check the below so it is in the same notation as Eq. (1). Also, should the $\delta\rightarrow \delta/2$?} \hlcy{RP: Looks like this notation matches. Do you think we need to restate the relationship between $p$ and $A$ and $B$? The $\delta_{AB}$  factor looks right to me because it should be $2$ when $A=B$ and $1$ when $A \neq B$.}

\begin{equation}
\begin{split}
    N_{A}N_{B} > \sqrt{\frac{2\epsilon_0}{m_{\rm e}}}\frac{(1+\delta_{AB}) N_{\rm min} q_e (E/n_{\rm gas})N_{\rm gas}}
    {\Delta \epsilon \min_p\left[\max_{\epsilon}{(c(\epsilon)\sigma_{p}(\epsilon))}\right]}, \\
    \label{NN_crit}
\end{split}
\end{equation}
where $\Delta \epsilon$ is the maximum change in energy per time step of a typical particle
with initial energy $\epsilon_0$. For simulations involving a fixed background gas, one may assume
some initial ionization percentage relating $N_A$ to $N_B$ and combine Eqs.~\eqref{NN_crit}, \eqref{Efield}
and \eqref{DTcrit} to solve for the initial number of particles, 
%\hl{MZ: What do you mean by field here? you mean in V/m? worth making a note of it} \hlcy{RP: Let me know if this is clear} 
the electric field (in V/m)
and the time step for a given
reduced field (in Td). For simulations involving a dynamic background gas (i.e. neutral heating or depletion),
one may use Eqs.~\eqref{DTcrit}, \eqref{Pcrit}, and \eqref{NTC} to solve for $N_A$, $N_B$, and $\Delta t$
for a given electric field value.

%\subsubsection{Sampling collision and acceleration processes}
%
%\textcolor{red}{\textbf{These are general ideas for this section but they need more formal statements and criteria:}}
%
%\textcolor{red}{(1) Collisions should occur often enough per time step that particles in the VDF do not unphysically run away. This generally means that the time spet needs to be large enough or that the number of particles great enough that Npairs is sufficiently large. The sign of runaway is generally an increasing mean energy. N.B. sometimes runaway is physical for large fields, one prominent example being coulomb collisions with an electric field above the the Dricer field.}
%
%\textcolor{red}{(2) Collisions should not occur so often per time step that the VDF begins to relax before an acceleration step.}
%
%\textcolor{red}{(3) Multiple collisions should be allowed each time step for very collisional systems. Physically, multiple collisions can occur before meaningful energy gain by an external field. The same should be true here. }
%
%\textcolor{red}{(4) Enough collisions should be present that each meaningful collision process can be sampled. Often the maximum time step is set by the cross section and a given maximum external field, independent of cell size. This means that the user needs to increase $N_aN_b/L^3$ while fixing $E/N$...  } 

\subsubsection{Stosszahlansatz (Molecular Chaos)}
As an aside, a notable benefit of the use of DSMC in 0D versus use in 1D, 2D, or 3D is that one can afford to use a large number of particles in a single cell. In large simulations meant to simulate realistic fluid flows it is often difficult to accommodate a large number of particles in a single cell due to the presence of thousands to billions of cells. Often, simulations may have as few as 10s of particles per cell. With so few particles it becomes difficult to satisfy Boltzmann's {\emph{Stosszahlansatz}} assumption, also known as the {\emph{molecular chaos}} assumption, that pre-collision particle velocities are uncorrelated. When the cells are sparsely populated the particle velocities of the selected colliding pair are correlated because a single prior collision significantly affects the phase space density used to evaluate the collision rate at the current step. It is well known that the {\emph{molecular chaos}} assumption is satisfied as the limit $N\to\infty$ is approached \cite{doi:10.1063/1.5099042}. In effect, this limit can be realized in 0D with the presence of thousands to hundreds of thousands of particles.

\subsection{Python API}
For additional user convenience 
we include a Python interface for ThunderBoltz. This will be especially
useful for users that wish to systematically manipulate cross section sets, input conditions, and any other simulation
settings made available within the ThunderBoltz input deck. The interface can read text file extracts from the LXCat database
\cite{https://doi.org/10.1002/ppap.201600098} and infer their configuration within ThunderBoltz. In python, one can easily plot,
scale, and generally manipulate cross section sets and procedurally prepare them for ThunderBoltz calculations.

For generating results at varying fields and densities, the interface provides optional implementation of
the input constraints imposed by Eqs.~\eqref{DTcrit},~\eqref{Efield}, and~\eqref{NN_crit}, as well as
automatic specification of reduced fields (i.e. $E/n_{\rm gas}$ or $B/n_{\rm gas}$). Based on specification
of $E/n_{\rm gas}$, $\epsilon_0$, $\Delta\epsilon$, and $\frac{N_{\rm e}}{N_{\rm gas}}$, the values
of $N_e$, $N_{\rm gas}$, $E$, and $\Delta t$ will be obtained automatically from the given cross section data set.

Once the input settings are configured, the interface will write the necessary ThunderBoltz input files,
compile, and run the program. During execution, there are options for live and asynchronous graphical monitoring of desired
transport parameters. 
After the simulation is finished, the interface can asynchronously read ThunderBoltz
output files and generate time-dependent and steady-state calculations for the electron mobility, mean energy,
reaction rate constants, and the Townsend ionization coefficient. Error bars for steady-state quantities
are determined by the standard deviation over several steady-state time steps. If velocity data are available,
one may easily produce joint and marginal velocity distribution plots, energy distribution plots, and quantile
statistics on the velocities and energies of the particles. Finally we provide examples using this Python API
which demonstrate these capabilities and reproduce the figures in this paper.

\section{Benchmarking Against Boltzmann Transport Calculations}

In this section, 
ThunderBoltz electron transport calculations are benchmarked against results from the two term Boltzmann solver Bolsig+\cite{Hagelaar_2005} and with Monte-Carlo benchmark data for electron transport in gases with crossed $E$ and $B$ fields\cite{White_1997}. 
The goal is to replicate the behavior of other established Boltzmann solvers on problems with well established solutions to demonstrate that the methods can be used interchangeably. This is an important first step that must be taken before applying the code to systems with a time-dependent background. In these tests, DSMC simulations are run while suppressing the change in species and temperature of the background population in chemical and excitative reactions. This is achieved by employing the \emph{Non-Reactive Collision} and \emph{Fixed Heavy Particle} collision models described in Sec. II. These options were first employed in a similar manner in the N\textsubscript{2} gas electron transport test problem in Ref.~\onlinecite{VVarticle} and are reproduced here for the case of Helium gas. Following these two tests, the ThunderBoltz chemical reaction functionality is benchmarked against an analytic chemical kinetic (reactive) problem, showcasing the utility of the code in the coupling of multiple reacting particle populations.    

\subsection{Ikuta-Sugai Benchmark Problem}

The Ikuta-Sugai Benchmark Problem has been used to test the calculation of electron transport in crossed electric and magnetic fields\cite{doi:10.1143/JPSJ.58.1228,White_1997}. The problem is specified as follows: A gas of particles with mass $m_{\rm gas}=100~m_e$ and number density of $n_{\rm gas}=10^{23}~\textrm{m}^{-3}$ has a z electric field component of $E_z/n_{\rm gas}=1$~Td. A tangential magnetic field of $B_y/n_{\rm gas}=$ 0, 10, 25, or 50~Hx (Hx is the unit Huxley and is defined as 1~Hx = $10^{-27}$~T~m$^3$) also applied. The gas has a single elastic electron-neutral interaction with a constant cross section of $\sigma_{el}=1.0~\mathring{\textrm{A}}^2$. 

\begin{figure}[hbt!]
\centering
\includegraphics[width=.5\textwidth]{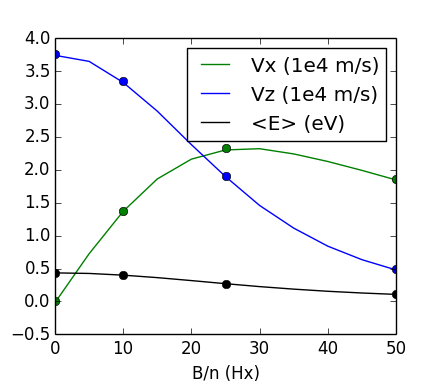}
\caption{\label{Fig:Ikuta} A comparison of ThunderBoltz DSMC simulation results (lines) with the data given by Ness\cite{Ness_1994} (dots) for the mean velocity and the x and z electron flow velocity as a function of magnetic field in the Ikuta-Sugai Benchmark Problem.}
\end{figure}

Previous authors have compared the values of diffusion, mean energy, and directional flow moments which are proportional to the mobility. For the present ThunderBoltz simulations, the electron flow components and mean energy are computed using Eqs.~(\ref{eq:vdirectional}) and (\ref{eq:meanE}). We omit the diffusion calculation as it has not yet been implemented in the present version of the code.
Figure.~\ref{Fig:Ikuta} shows the comparison of the calculated values with the published results from Ref.~\onlinecite{Ness_1994}. The values are in agreement with the the know values within the extent demonstrated by other codes\cite{White_1997}.
           
%\subsection{Reid Ramp Model}

%\subsection{Electron Reaction Rates and Swarm Parameters in Helium\label{sec:HeSwarm}}
\subsection{Electron Rate Constants and Swarm Parameters in Helium\label{sec:HeSwarm}}

In this test problem, Helium %reaction rates 
rate constants
and transport coefficients were calculated using
the approach outlined in section~\ref{conv_cons} and the results were benchmarked against results
from Bolsig+ and swarm experiment data. 
%\hlcy{BSS: Ionization fraction is unimportant since you are not modeling the background.} \hl{RP: Okay, that makes sense. Its also approaching $\infty$ when 'egen' is on anyway. I include it here because it is a required parameter when using the 'autostep' option} 
Since the background in these simulations is fixed the ionization fraction is unimportant and additional electrons can be utilized to improve statistics. The simulation used an initial ionization ratio of $\frac{N_{\rm
e}}{N_{\rm gas}}~=~10$, a minimum acceleration energy of $\Delta \epsilon~=~0.1$~eV, and an
initial energy of $\epsilon_0~=~$10~eV. The time step $\Delta t$, and number of particles,
$N_{\rm e}$, $N_{\rm He}$, were determined by Eqs.~\eqref{DTcrit},~\eqref{Efield}, and~\eqref{NN_crit}
at each reduced field. The cell length was fixed at $L=1~\mu{\rm m}$.
Cross sections were prepared from a combination of tabulated CCC~\cite{bray2011} data and analytic fits. In order to preserve near-threshold behavior, the  tabular CCC data was used below
300~eV, and the analytic fits  were used above 300~eV for all processes. 
The elastic model includes a momentum transfer cross section obtained by
integrating the differential cross section forms in
Ref.~\onlinecite{Park_2022} which derive from CCC data. Analytic fits for ten excitation cross
sections ($n \leq 3$), and one ionization
cross section were employed from analytic fits to CCC cross sections provided by Ref.~\onlinecite{Ralchenko2008}. Isotropic scattering is assumed for all collision models.
The ionization model was configured to track ejected electrons, so each ionization
event increases $N_{\rm e}$ by 1. The reaction rates were calculated using
Eq.~(\ref{eq:rates}), while the bulk and flux calculations for the electron mobility and Townsend
ionization coefficients were obtained from Eqs.~(\ref{eq:vflux}), (\ref{eq:bflux}), (\ref{eq:mobility}),
and (\ref{eq:alpha}). Simulations were carried out using both the equal and one-takes-all electron
energy sharing models for electron impact ionization.

\begin{table}[p!]
\caption{\label{tab:ratecomp}  Comparison of  rate constants calculated from Bolsig+ and ThunderBoltz for 50~Td and 500~Td cases. The \emph{one takes all} ionization energy sharing model was assumed.
Bolsig+ calculations were performed with the 
gradient expansion (GE) and steady-state Townsend (SST) growth models.
%\hl{MZ: This is actually referred to as steady-state Townsend or spatial growth. Are we referring to this correctly?} \hlcy{RP: See above where I've made the same change.}
}
\centering
\begin{tabular}{lccccc}
\hline
Quantity & ThunderBoltz 50~Td  & Bolsig GE 50~Td & Bolsig SST 50~Td & ThunderBoltz 500~Td & Bolsig GE 500~Td \\
\hline
$k_1$ (m$^3$/s) &7.67(0.001)E-14&7.69E-14&7.70E-14&4.49(0.02)E-14&4.80E-14\\
$k_2$     &4.18(0.02)E-17&4.20E-17&3.88E-17&3.81(0.03)E-16&4.17E-16\\
$k_3$     &3.43(0.02)E-17&3.41E-17&3.08E-17&2.09(0.01)E-15&2.23E-15\\
$k_4$     &6.22(0.03)E-17&6.27E-17&5.86E-17&1.64(0.01)E-16&1.80E-16\\
$k_5$     &2.19(0.05)E-18&2.20E-18&2.00E-18&3.27(0.03)E-17&3.62E-17\\
$k_6$     &4.66(0.07)E-18&4.66E-18&4.24E-18&7.49(0.06)E-17&8.21E-17\\
$k_7$     &3.60(0.02)E-17&3.62E-17&3.33E-17&1.87(0.01)E-16&2.08E-16\\
$k_8$     &1.53(0.04)E-18&1.55E-18&1.42E-18&6.81(0.09)E-18&7.50E-18\\
$k_9$     &7.80(0.09)E-18&7.88E-18&7.24E-18&3.53(0.03)E-17&3.91E-17\\
$k_{10}$  &5.82(0.08)E-18&5.77E-18&5.16E-18&4.92(0.03)E-16&5.23E-16\\
$k_{11}$  &5.61(0.08)E-18&5.61E-18&5.09E-18&4.48(0.04)E-17&5.00E-17\\
$k_{12}$  &4.39(0.03)E-17&4.31E-17&3.77E-17&7.32(0.05)E-15&7.73E-15\\
%$\mu_eN$ (V/(m s))&1.068e+24&[Value]& 7.702e+23&[Value]&6.28[.23]e+23\\
\hline
\end{tabular}
\end{table}
\begin{figure}[hbt!]
\centering
\includegraphics[width=.5\textwidth]{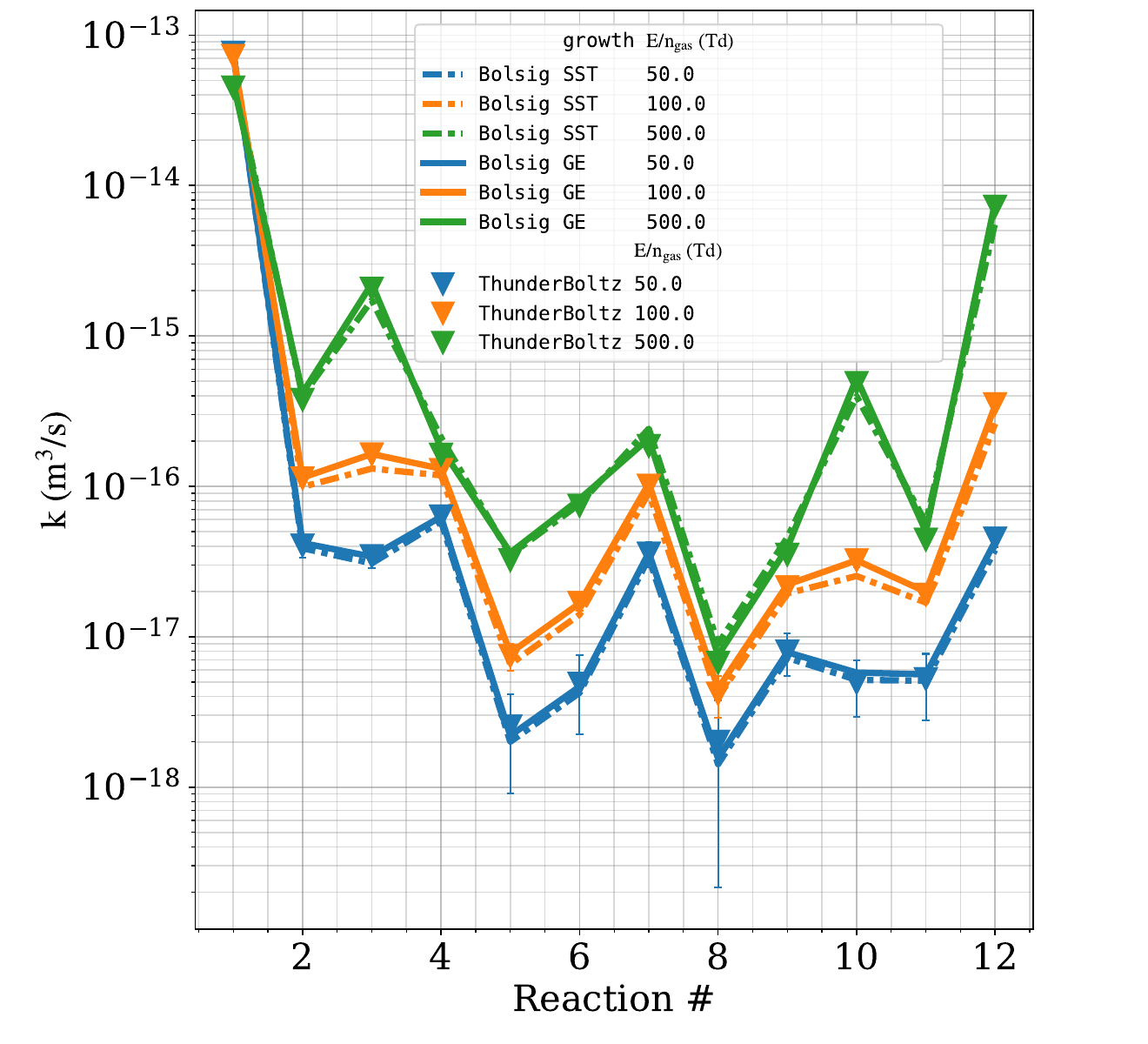}
\caption{A comparison of the electron rate constants in He from ThunderBoltz and Bolsig+ for reduced electric fields of 50, 100, and 500~Td. The \emph{one takes all} ionization energy sharing model was assumed.}
\label{fig:ratecomp}
\end{figure}

In Fig.~\ref{fig:ratecomp} and Table~\ref{tab:ratecomp} the rate constants for the elastic scattering ($k_1$), ten inelastic loss processes($k_2$--$k_{11}$), and ionization ($k_{12}$) are compared with Bolsig+ calculations using the gradient expansion (GE) and steady-state Townsend (SST) 
%\hl{MZ: This is actually referred to as steady-state Townsend or spatial growth. Are we referring to this correctly?} \hlcy{RP: I think we should use steady-state Townshend, since that is what SST stands for in Bolsig (even though it is steady-state transport). I've changed ``methods" to ``growth models" to emphasize that it is a model corresponding to the ``Steady State Townsend experiments" as is says in the manual.}
growth models (See Ref.~\onlinecite{Hagelaar_2005} section 2.2). The Bolsig+ data calculated with the GE method is in good agreement with the ThunderBoltz results for $E/N=50$~Td, while small discrepancies exist for the SST data. This is expected since the SST method assumes a flux of electrons from the cathode that grows exponentially in space. Since this source flux is not present in the ThunderBoltz simulations the results are expected to differ from the SST calculations. This difference along with the convergence criteria appears to resolve a previous discrepancy between rate constants from particle and Bolsig+ calculations that was previously discussed in Ref.~\onlinecite{VVarticle}. At larger values of $E/N= 500$~Td the values differ slightly as shown in Table~\ref{tab:ratecomp} and is expected for higher field values where the two term approximation breaks down\cite{Stephens_2018}.   

\begin{figure}[hbt!]
%\begin{figure}[H]
\centering
\includegraphics[width=.5\textwidth]{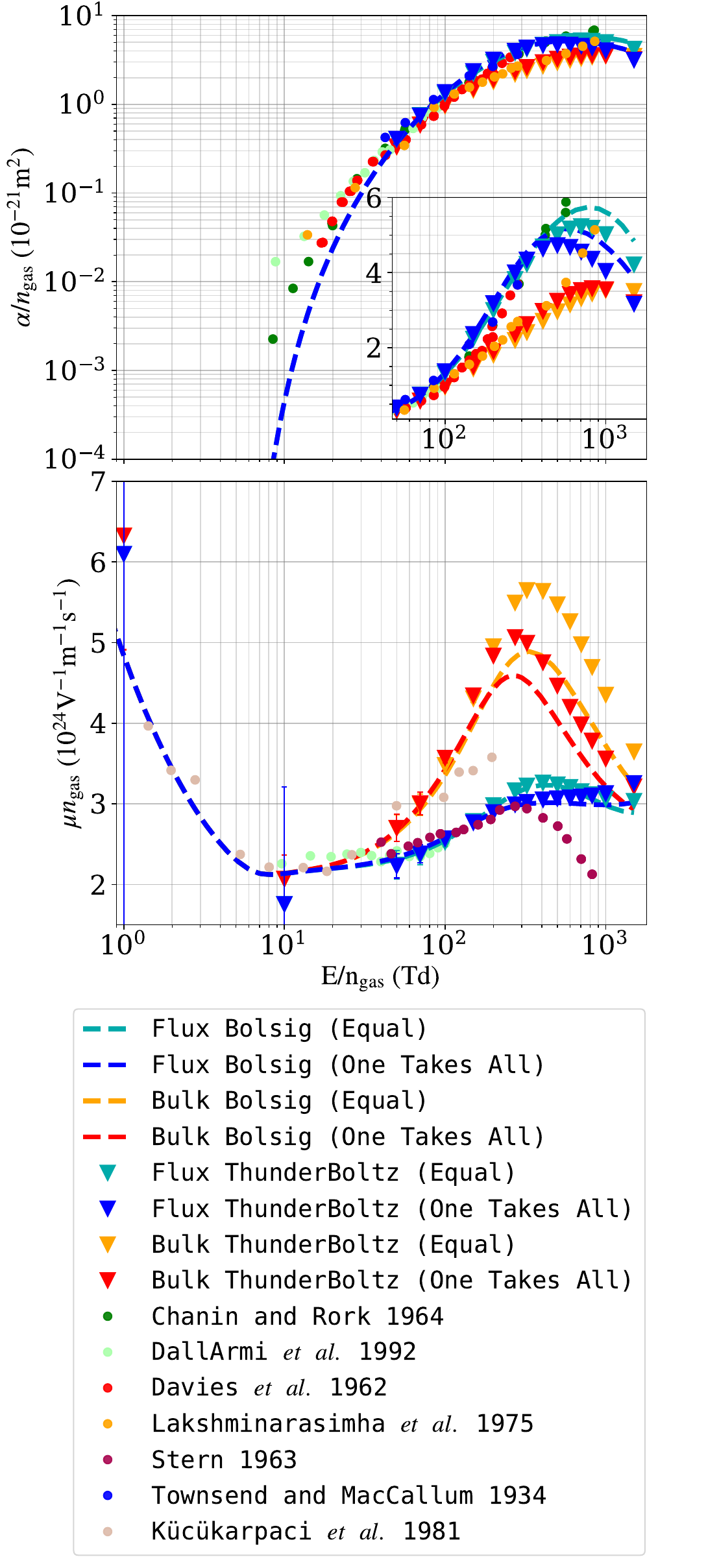}
\caption{\label{fig:TBtransport}The reduced Townsend ionization coefficient (top), and reduced electron mobility (bottom) at various
reduced fields. Bulk values are defined from the swarm drift velocity in the negative field direction, whereas the flux values
are defined from the first EVDF moment in the negative field direction. The `Equal' and `One Takes All' parameters refer
to the assumed electron energy sharing ionization model. Isotropic scattering is assumed. Experimental data (dots) are obtained from the LXCat
website.\cite{https://doi.org/10.1002/ppap.201600098}}
\end{figure}

Figure~\ref{fig:TBtransport} compares ThunderBoltz calculations of flux and bulk coefficients with Bolsig+ and experimental measurement, obtained from the LXCat database\cite{https://doi.org/10.1002/ppap.201600098}. For the Bolsig+ comparison, the flux and bulk coefficients are in good agreement for both equal and one-takes-all energy sharing models, with the expected deviations at larger $E/N$. The comparison between measured data to computed flux and bulk coefficients requires care due to the variety of swarm experiment configurations used to obtain the data, whether the particular configuration probes transport in the hydrodynamic or non-hydrodynamic regimes, and whether or not the experiment measures values in the swarm center of mass frame\cite{PT-Casey_2021,Swarm-Petrovic_2009,1984AuJPh..37..593B}. Generally, the measured data is in good agreement with the simulation results for bulk or flux coefficients, where appropriate, but is not constraining enough to distinguish different electron energy sharing models at high values of $E/N$.
\iffalse
but is not constraining enough to distinguish different electron energy sharing models.
since differences between models are large at high values of $E/N$, where measured data currently does not exist or exhibit a large spread. 
\fi
 The details of the data comparison is summarized in the following paragraph.

For the non-hydrodynamic regime, steady-state Townsend experiments probe the flux ionization coefficients. As expected, the data from SST experiments of Townsend and MacCallum\cite{doi:10.1080/14786443409462426},
Chanin and Rork\cite{PhysRev.133.1005}, and Davies~\etal\cite{DKennethDavies_1963} agree with the Thunderboltz and Bolsig+ flux calculations over the range of 50--600~Td.

In the hydrodynamic regime the comparison is further complicated by the fact that a large portion of the experiment data was collected using the pulsed Townsend (PT) method, which in most cases  extracted coefficients are based on analyzing the current collection using the Brambring diffusion equation, which cannot be derived from the Boltzmann equation\cite{PT-Casey_2021}. These coefficients are neither flux nor bulk coefficients, but can be converted in some cases. In the case where the volume ionization source is unimportant, these correspond to the flux coefficient values. In several measurements for both Ar and SF$_6$ gases, the coefficients are close to those of the flux values at higher $E/N$ (see Figs. 2 and 3 in Ref. \onlinecite{PT-Casey_2021}). Curiously, the PT data of Dall'Amari~\etal\cite{1992AuJPh..45..185D} agrees with the flux mobility up to 100~Td. 
%\hl{Is the PT configuration expected to be in the hydrodynamic regime or agree with the bulk data? BSS: It is in the hydrodynamic regime, but much (or possibly all of the data?) was interpreted from a form of the diffusion equation that cannot be derived from the Boltzmann equation so the form of transport is inconsistent. See ref PT-Casey2021 from the paragraph above.}
This could be either due to a reduced importance of the bulk ionization, or the utilization of methods to extract flux values using more detailed analysis realized in an earlier paper\cite{1984AuJPh..37..593B}. Also probing the hydrodynamic regime are the time of flight experiments of K{\"u}c{\"u}karpaci~\etal\cite{HNKucukarpaci_1981} and Lakshminarasimha~\etal \cite{piee.1975.0287}. These agree with the bulk mobility and ionization coefficients up to 100~Td and 500~Td, respectively. The methods of these two data sets are expected to measure the bulk transport coefficients, consistent with the agreement between data and simulation. Finally, the data of Stern\cite{} is found to be in good agreement with the flux predictions, although the experimental configuration could not be determined due to an inability to locate the text of the original reference.

%Mobility:
%Takeda ~\etal 1982 
%In agreement below 50Td (3 TB data points)

\subsection{Onsager Relation for Coupled Chemical Reactions\label{sec:onsager}}
This test aims to reproduce the equilibrium condition 
%\hl{MZ: I changed the density index in the below equation in order to match the (17--22) notation}
%\hlcy{RP: I'm flipping the RHS and LHS just so $i$ comes before $j$.}
\begin{equation}\label{eq:onsager}
%    n_i k_{ji}=n_jk_{ij}
%    n_j k_{ji}=n_i k_{ij}
     n_i k_{ij}=n_j k_{ji}
\end{equation}
by considering a realization of the generic set of interactions $A \rightleftarrows B \rightleftarrows C \rightleftarrows A$ used to demonstrate a chemical reaction analog of the reciprocal relations in Onsager's famous 1931 paper\cite{PhysRev.37.405}. Like the reciprocal relations for transport phenomena, the relation Eq.~\eqref{eq:onsager} follows from the 
%microscopic reversibility 
%\hl{MZ: changed from ``microscopic reversibility" to the principle of detailed balance. Detailed balance refers to equating rates, while microscopic reversibility refers to cross section relations.}
the principle of detailed balance
of the collision processes and the positive entropy generation for processes that result in fluctuations around equilibrium.

In this realization of Onsager's analogy, particle species $A$, $B$, and $C$  are populated or depopulated via collisions with a background species $X$. The collisional processes are described by the following cross sections
\begin{equation}
    A+X\to B+X \ \ \ \sigma_{AB}(\epsilon)=H(\epsilon-1.0~\textrm{eV})~\overcirc{\textrm{A}}^2,
\end{equation}
\begin{equation}
    B+X\to C+X \ \ \ \sigma_{BC}(\epsilon)=2 \: H(\epsilon-1.0~\textrm{eV})~\overcirc{\textrm{A}}^2,
\end{equation}
\begin{equation}
    A+X\to C+X \ \ \ \sigma_{AC}(\epsilon)=2 \: H(\epsilon-1.0~\textrm{eV})~\overcirc{\textrm{A}}^2,
\end{equation}

\begin{equation}
    B+X\to A+X \ \ \ \sigma_{BA}(\epsilon)=1~\overcirc{\textrm{A}}^2,
\end{equation}
\begin{equation}
    C+X\to B+X \ \ \ \sigma_{CB}(\epsilon)=2~\overcirc{\textrm{A}}^2,
\end{equation}
\begin{equation}
    C+X\to A+X \ \ \ \sigma_{CA}(\epsilon)=3~\overcirc{\textrm{A}}^2,
\end{equation}
where $H(\epsilon)$ is a heavy-side step-function.
The benefit of the above cross sections is that the rate constants are known analytically and can be compared with those calculated from the simulation. For the cross section 
\begin{equation}\label{eq:hs}
\sigma(\epsilon)=    \begin{cases}
      0, &  \epsilon<E_a \\
      \pi d^2, &  \epsilon\ge E_a
    \end{cases}
\end{equation}
the rate constant is
\begin{equation}\label{eq:arrRate}
k(T)= 2d^2\bigg(\frac{2\pi k_B T}{m_r}\bigg)^{1/2}\exp\bigg(\frac
{-E_a}{k_B T}\bigg)\bigg(1+\frac
{E_a}{k_B T}\bigg).
\end{equation}
In addition to these, the species $A$, $B$, and $C$ collide with themselves 
and the background, $X$, with a
constant cross section of $100~\overcirc{\textrm{A}}^2$, and the background collides with itself with a constant cross section of $1~\overcirc{\textrm{A}}^2$
%\hl{RP: I believe this should be $1~\overcirc{\textrm{A}}^2$ for the X+X reactions and $100~\overcirc{\textrm{A}}^2$ for all the other non reacting collisions (to match the figure / input files)}\hlcy{BSS: Ryan, go ahead and change it to whatever you have for your input files.} \hl{RP: Done} 
to ensure that the
species remain at a common temperature. Initially, the background, $X$, is at a density of $2\times10^{24}$~m$^{-3}$
and species $A$ is at $1\times10^{22}$~m$^{-3}$, with $B$ and $C$ at zero.

%\begin{figure}[hbt!]
%\centering
%\includegraphics[width=.5\textwidth]{TestFig.png}
%\caption{\hlcy{RP: proposing this caption:} The time-dependent density (top), reaction rate constant (middle), and absolute reaction rates (bottom) for the Onsager system. The rate constants produced by ThunderBoltz are shown to converge to the analytic values of Eq.~\eqref{eq:arrRate} (solid horizontal lines).} \hl{POPULATE} \hlcy{RP: Do we want to include a legend for the middle panel?}\hl{BSS: Ryan, did you have an updated figure for this, or did you want me to change this one?} \hlcy{I do have  one, which I'll include below, but also happy with the one above.
%\hl{MZ: Change bottom panel  y-axis label and legend to lower case 'k' and middle panel 'Rate' to 'k'.}}
%\end{figure}

\begin{figure}[hbt!]
\centering
\includegraphics[width=.5\textwidth]{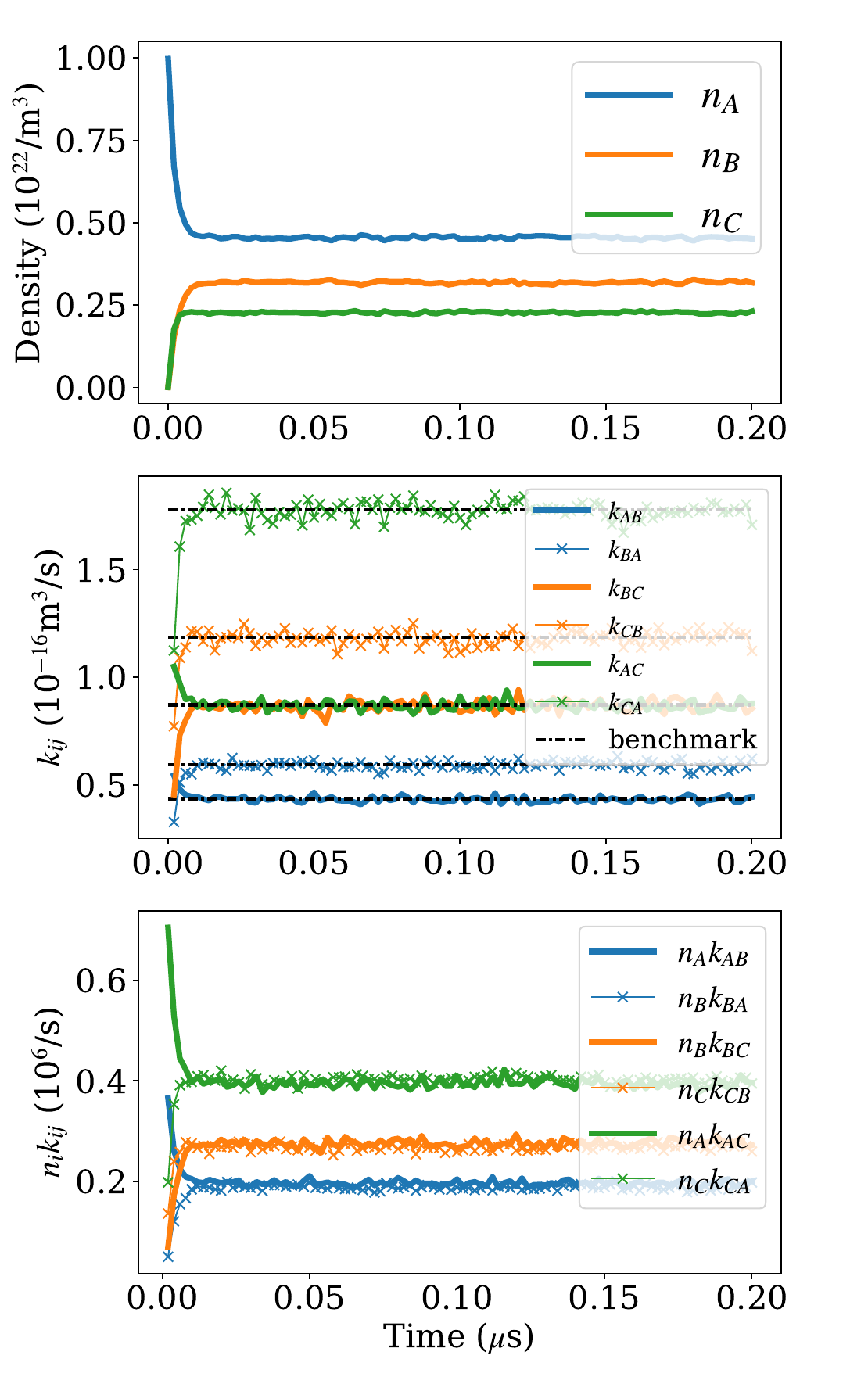}
\caption{The time-dependent density (top), reaction rate constant (middle), and absolute reaction rates (bottom) for the Onsager system. In the middle panel the rate constants produced by ThunderBoltz are shown to converge to the analytic values of Eq.~\eqref{eq:arrRate} (dash-dotted horizontal lines).
}
\end{figure}

\section{Applications}
\subsection{Gas Heating and Ion Transport\label{sec:gasHeating}}
One benefit of the use of a 0D DSMC simulation over other existing methods is that it is capable of calculating electron and ion transport for systems where time-dependence and non-stationarity of the background is important due to its interaction with charged species. One such situation is the heating of a neutral gas due to current flow where the ionization fraction of the gas is great enough for the charge-neutral collisions to contribute to direct heating of the gas. The neutral and ion temperatures are initially out of equilibrium. As the simulation proceeds, charged particles are accelerated in the fixed field and collide with neutrals, heating the background. 
Due to the large mass difference between electrons and neutrals these collisions are ineffective at transferring energy and nearly all direct collisional heating occurs from ion-neutral collisions. 
Such situations likely resemble elements of the physics present in electrostatic discharges where the ionization fraction can be of the order of $\tildetext$10\%.
%\cite{?} \hl{MZ references}\hlcy{BSS: I looked around for 40 minutes and I can't seem to find a reference for this. I believe that I confirmed this with the CSM guys.}. 
However, in these discharges, quenching of electronic states also plays a significant role in determining the gas temperature on sub-microsecond timescales\cite{Popov_2011,Popov_2016}. 

\begin{figure}[hbt!]
\centering
\includegraphics[width=.5\textwidth]{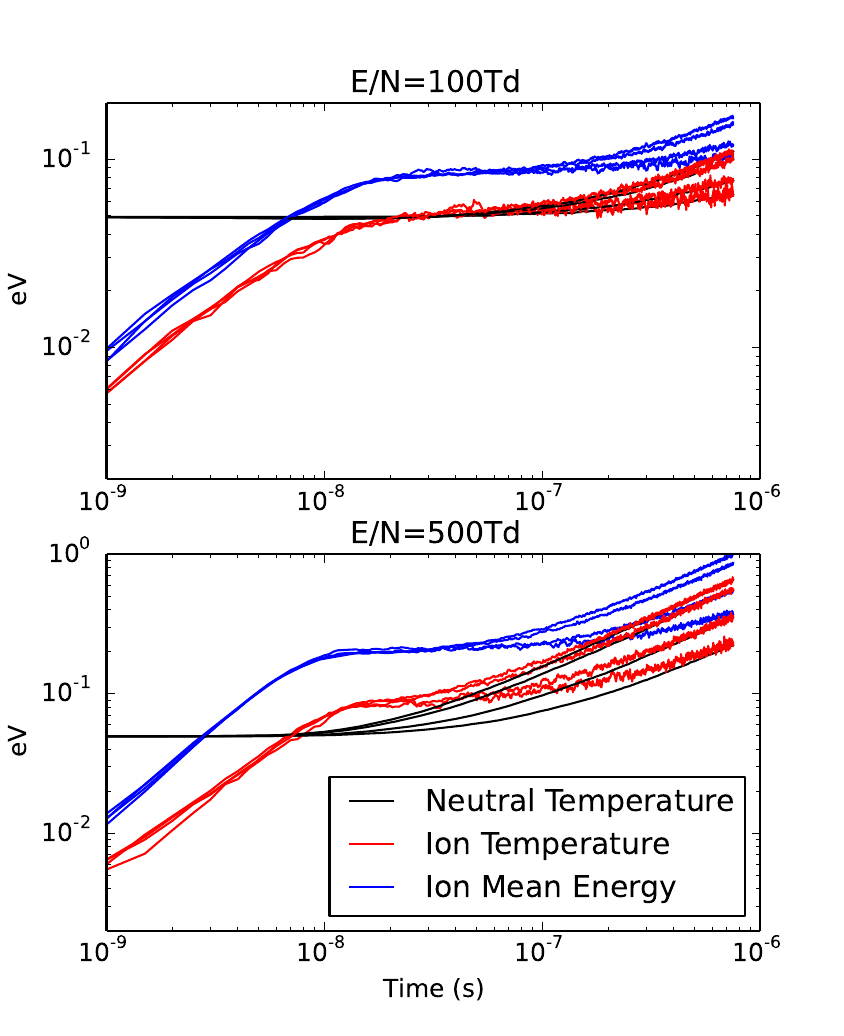}
\caption{\label{Fig:ionheating}The ion and neutral temperatures and ion mean energy for $E/N = 100$~Td (top) and 500~Td (bottom). In each grouping of lines, the results for 1\%, 2\%, 4\%, and 5\% ionization are shown in ascending order of the magnitude. Heating rates occur faster for greater ionization fraction. 
%\hl{MZ: Hard to see the top-panel neutral temperature, perhaps plot last?}
}
\end{figure}

\begin{figure}[hbt!]
\centering
\includegraphics[width=.5\textwidth]{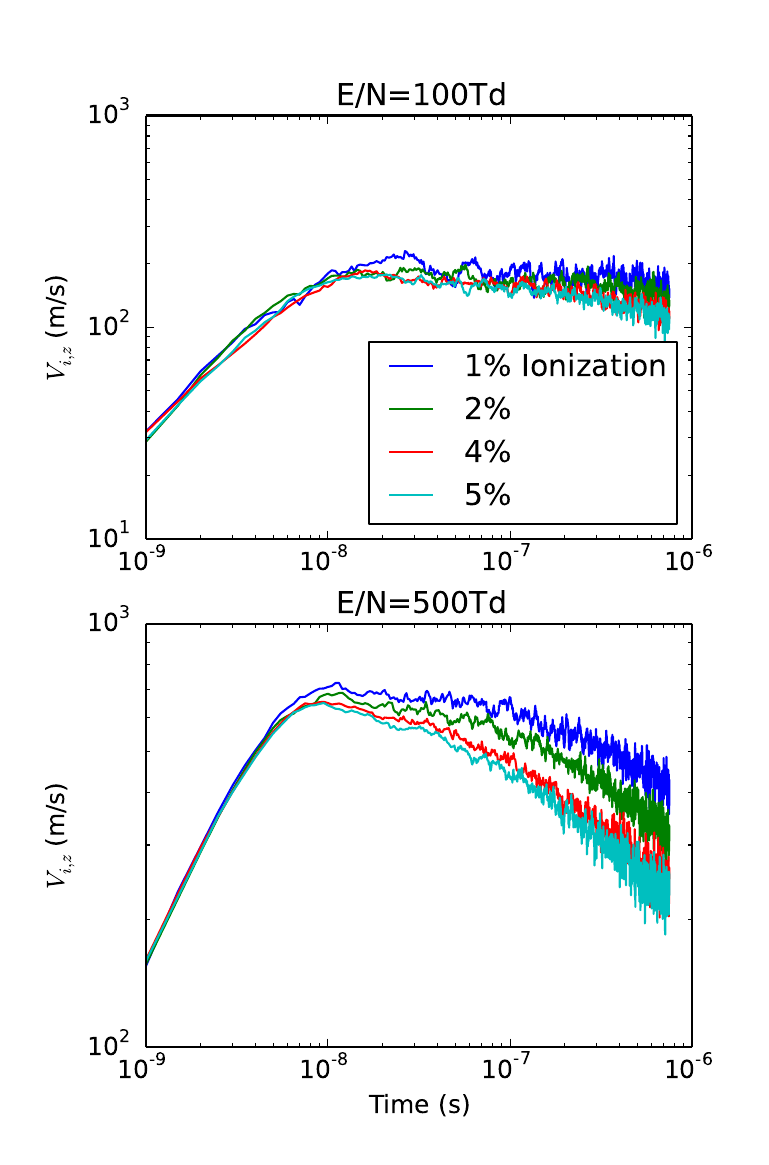}
\caption{\label{Fig:ionflow}The ion flow velocity as a function of time for 1\%, 2\%, 4\%, and 5\% ionization of N\textsubscript{2} gas at $E/N = 100$~Td (Top) and 500~Td (bottom). }
\end{figure}

In this section we demonstrate the modeling of the ion-neutral heating of N\textsubscript{2} gas. 
The gas model, which is summarized in Table ~\ref{tab:tableN2}, includes ions and electrons at a fixed ionization fraction in a background of N\textsubscript{2} by treating the electron-impact ionization cross section as an inelastic loss process with the \emph{Non-Reactive Collision} model. Electron elastic energy transfer and inelastic losses due to excitation and ionization are modeled using the N\textsubscript{2} cross sections in the TRINITI database\cite{TRINITI} obtained from LXCat\cite{https://doi.org/10.1002/ppap.201600098}. In addition to these, ion-neutral and neutral-neutral interactions are modeled using a hard sphere model based on the determined atomic radii. Simulations are initialized with all particles at rest, a set fractional ionization of 1\%, 2\%, 4\%, or 5\% ionization in nitrogen gas at a density of $n_{N_2}=10^{23}\textrm{m}^{-3}$ of nitrogen and $E/N = 100$~Td or 500~Td.
The simulated current-driven heating of the gas is shown in the two panels of Fig.~\ref{Fig:ionheating}, which displays the ion and neutral temperature and ion mean energy for each ionization fraction. Corresponding ion flow velocities are shown in Fig.~\ref{Fig:ionflow}.

\begin{widetext}

\begin{table}[p!]
\caption{\label{tab:tableN2} N\textsubscript{2} Gas Model}
\centering
\begin{tabular}{lcccc}
\hline
Number & Reaction & Simulated Reaction & $\Delta E$ (eV) & Ref.[]\\
\hline
1&e + N$_2$ $\to$ e + N$_2$&e + N$_2$ $\to$ e + N$_2$&-&\onlinecite{TRINITI} \\
2&e + N$_2$ $\to$ e + N$_2(\nu=1)$&e + N$_2$ $\to$ e + N$_2$&0.29&\onlinecite{TRINITI} \\
3&e + N$_2$ $\to$ e + N$_2(\nu=2)$&e + N$_2$ $\to$ e + N$_2$&0.59 &\onlinecite{TRINITI}\\
4&e + N$_2$ $\to$ e + N$_2(\nu=3)$&e + N$_2$ $\to$ e + N$_2$&0.88&\onlinecite{TRINITI} \\
5&e + N$_2$ $\to$ e + N$_2(\nu=4)$&e + N$_2$ $\to$ e + N$_2$&1.17 &\onlinecite{TRINITI}\\
6&e + N$_2$ $\to$ e + N$_2(\nu=5)$&e + N$_2$ $\to$ e + N$_2$&1.47 &\onlinecite{TRINITI}\\
7&e + N$_2$ $\to$ e + N$_2(\nu=6)$&e + N$_2$ $\to$ e + N$_2$&1.76&\onlinecite{TRINITI} \\
8&e + N$_2$ $\to$ e + N$_2(\nu=7)$&e + N$_2$ $\to$ e + N$_2$&2.06&\onlinecite{TRINITI} \\
9&e + N$_2$ $\to$ e + N$_2(\nu=8)$&e + N$_2$ $\to$ e + N$_2$&2.35&\onlinecite{TRINITI} \\
10&e + N$_2$ $\to$ e + N$_2$ A$^3\Sigma (\nu=0-4)$&e + N$_2$ $\to$ e + N$_2$&6.17&\onlinecite{TRINITI} \\
11&e + N$_2$ $\to$ e + N$_2$ A$^3\Sigma (\nu=5-9)$&e + N$_2$ $\to$ e + N$_2$&7.00 &\onlinecite{TRINITI}\\
12&e + N$_2$ $\to$ e + N$_2$ B$^3\Pi$&e + N$_2$ $\to$ e + N$_2$&7.35&\onlinecite{TRINITI} \\
13&e + N$_2$ $\to$ e + N$_2$ W$^3\Delta$&e + N$_2$ $\to$ e + N$_2$&7.36&\onlinecite{TRINITI} \\
14&e + N$_2$ $\to$ e + N$_2$ A$^3\Sigma (\nu=10-)$&e + N$_2$ $\to$ e + N$_2$&7.80&\onlinecite{TRINITI} \\
15&e + N$_2$ $\to$ e + N$_2$ B$^{'3}\Sigma$&e + N$_2$ $\to$ e + N$_2$&8.16 &\onlinecite{TRINITI}\\
16&e + N$_2$ $\to$ e + N$_2$ a$^{'1}\Sigma$&e + N$_2$ $\to$ e + N$_2$&8.40 &\onlinecite{TRINITI}\\
17&e + N$_2$ $\to$ e + N$_2$ a$^1\Pi$&e + N$_2$ $\to$ e + N$_2$&8.55 &\onlinecite{TRINITI}\\
18&e + N$_2$ $\to$ e + N$_2$ w$^1\Delta$&e + N$_2$ $\to$ e + N$_2$&8.89&\onlinecite{TRINITI} \\
19&e + N$_2$ $\to$ e + N$_2$ C$^3\Pi$ &e + N$_2$ $\to$ e + N$_2$&11.03 &\onlinecite{TRINITI}\\
20&e + N$_2$ $\to$ e + N$_2$ E$^3\Sigma$&e + N$_2$ $\to$ e + N$_2$&11.88&\onlinecite{TRINITI} \\
21&e + N$_2$ $\to$ e + N$_2$ a$^{''1}\Sigma$&e + N$_2$ $\to$ e + N$_2$&12.25&\onlinecite{TRINITI}\\
22&e + N$_2$ $\to$ e + N$_2$ (sum of singlets) &e + N$_2$ $\to$ e + N$_2$&13.0 &\onlinecite{TRINITI}\\
23&e + N$_2$ $\to$ e + e + N$_2^+$&e + N$_2$ $\to$ e + N$_2$&15.6&\onlinecite{TRINITI} \\
24&N$_2$ + N$_2$ $\to$ N$_2$ + N$_2$&N$_2$ + N$_2$ $\to$ N$_2$ + N$_2$&- &Hard Sphere Diameter: \cite{bird1994molecular}\\
25&N$_2^+$ + N$_2$ $\to$ N$_2^+$ + N$_2$&N$_2^+$ + N$_2$ $\to$ N$_2^+$ + N$_2$&-&Hard Sphere Diameter: \cite{bird1994molecular} \\
\hline
\end{tabular}
\end{table}

\end{widetext}

\begin{table}[p!]
\caption{\label{tab:comparBolsig}  Comparison of results from Bolsig+ and ThunderBoltz. The standard deviation of time dependent fluctuations is given in brackets (e.g. .004 in 1.159[.004]E-13).  }
\centering
\begin{tabular}{lcc}
\hline
Quantity & Bolsig 500Td &  500Td w/ Heating \\
&&($\Delta t=5\cdot10^{-12} s)$\\
\hline
$k_1$ (m$^3$/s)&1.937E-13&1.932[.008]E-13\\
$k_2$&2.442E-15&2.369[.100]E-15\\
$k_3$&1.138E-15&1.087[.063]E-15\\
$k_4$&7.639E-16&7.257[.580]E-16\\
$k_5$&5.243E-16&4.852[.456]E-16\\
$k_6$&4.415E-16&4.055[.389]E-16\\
$k_7$&3.751E-16&3.473[.370]E-16\\
$k_8$&2.063E-16&1.878[.299]E-16\\
$k_9$&9.922E-17&8.324[1.84]E-17\\
$k_{10}$&1.206E-16&$^a$\\
$k_{11}$&5.052E-16&3.830[.305]E-16\\
$k_{12}$&1.535E-15&1.459[.063]E-15\\
$k_{13}$&1.592E-15&1.560[.078]E-15\\
$k_{14}$&4.873E-16&4.789[.354]E-16\\
$k_{15}$&4.694E-16&3.999[.344]E-16\\
$k_{16}$&3.689E-16&2.970[.316]E-16\\
$k_{17}$&1.306E-15&1.250[.059]E-15\\
$k_{18}$&4.425E-16&4.160[.377]E-16\\
$k_{19}$&3.409E-15&3.397[.098]E-15\\
$k_{20}$&3.967E-17&3.456[1.11]E-17\\
$k_{21}$ &1.535E-16&1.234[.202]E-16\\
$k_{22}$& 3.82E-15&3.755[.117]E-15\\
$k_{23}$ &1.801E-15&1.712[.079]E-15\\
$\langle\mathcal{E}\rangle$ (eV)&9.443&9.39[.078]\\
\hline
\end{tabular}
\

\small $^a$For this cross section no reactions were recorded. Reaction 10 has the smallest $(c\sigma)_{max}$ and due to the order of magnitude smaller time step compared to simulations without heating evaluation of $N_{pairs}$ results in a quantity that rounds to zero when converting to an integer.   
\end{table}

As shown in Figs.~\ref{Fig:ionheating} and \ref{Fig:ionflow}, each simulation exhibits an initial period at $t<10^{-8}$~s where the ion temperature, mean energy, and flow velocity increases with roughly a power-law behavior until reaching a plateau. During this time, the behavior is independent of ionization fraction. This behavior is due to the establishment of an initial \emph{quasi-equilibrium} as the ions accelerate from their initial condition at rest. On this time scale, as indicated by the neutral temperature, there have been insufficient collisions with neutrals to appreciably change the neutral temperature and the behavior is similar to what one would expect for ions flowing against a fixed background. 

Initially after the \emph{quasi-equilibrium} is established, the ion and neutral temperatures are out of equilibrium.
As time progresses, ions gradually heat the neutrals, eventually coming into equilibrium with them. Figure~\ref{Fig:ionheating} shows that the time required for this to occur increases with increasing electric field strength. 
It is also apparent from comparing the 100 and 500~Td simulations that the time scale for ion and neutral temperatures to equilibrate decreases with increasing ionization fraction. This can be understood as a result of having a greater number of ions available to heat the neutral gas. The interaction between ions and neutrals strongly affects the properties of each species creating different time-dependent behavior for each ionization fraction. 
This difference is expected to affect ion transport coefficients as well. 
For example, the ion flow velocity in Fig.~\ref{Fig:ionflow} differs significantly for simulations with different ionization fractions. 
For hard sphere collisions, the larger gas temperature facilitated by the increased ions-neutral collisions at larger ionization fraction is expected to further increase the collision rate due to the larger value of the characteristic relative velocity of a particle pair $c$ and the value of $\left[c\sigma_p\right]_{\rm max}$. This increased collisionality decreases the ion flow and hence the ion mobility which is defined by the phenomenological relation $\textbf{E}=\mu_s\textbf{V}_s$.

Unlike the ion-neutral behavior, heating of the neutrals needs to be significant for it to have an effect on the electron transport properties because the typical electron energy is so much greater than that of the neutral gas.
Even when the gas is heated to a tenth of an eV, an appreciable amount for applications that aim to modify gas flow properties\cite{Leonov_2016}, the center of mass energy of a typical electron-neutral collision changes very little. 
This feature of electron behavior is demonstrated in Table~\ref{tab:comparBolsig},
which shows the reaction rates obtained at the end of the simulation for the 500~Td 5\% ionization simulation still agree with Bolsig+ results that do not account for ion-neutral background heating. This demonstrates that the electron-neutral collision rates remain nearly constant while the rates for neutral-neutral and ion-neutral (Table~\ref{tab:tableN2} Reactions 24 and 25) increase due to the heating of the gas. However, in some situations one may expect situations to arise where this is not the case. For example, with sufficient electron and ion densities, collisions between electrons and ions may affect how each charged particle species couples to the neutral gas.

%\begin{figure}[hbt!]
%\centering
%\includegraphics[width=.5\textwidth]{RatesHeating.pdf}
%\caption{\label{Fig:ratesheating}}
%\end{figure}

These results indicate that significant gas heating can occur in discharges with large ionization fractions. Even at the most modest conditions explored of 100~Td and 1\% ionization the background gas increases in temperature by 100~K in less than 1 microsecond, a timescale that is orders of magnitude shorter than heating from V-T transfer. These results suggest that ion-neutral heating may play a significant role for discharges that are in operation over long durations and that further exploration comparing ion heating rates to those from electronic state quenching and V-T transfer is warranted in more detailed gas models than the one presented here.

\subsection{Plasma Kinetics Simulations\label{sec:kinetics}}

%As a final example, a plasma chemical kinetics type simulation is considered. Here, the time dependent densities of the first two nitrogen vibrational excitations are considered in a discharge driven by sinusoidal electric field. 
%As a final example, a
A plasma chemical kinetics type simulation is considered here, where the time dependent densities of the first two N\textsubscript{2} vibrational excitations are considered in a discharge driven by a sinusoidal electric field. 
To track the first two excitations, the reactions 1 and 4--23 of Table~\ref{tab:tableN2} 
are supplemented by the reactions F1--F8 of Table~\ref{tab:table2}. The reactions F1--F6 of Table~\ref{tab:table2} are from the Flinders database\cite{FLINDERS}, described in Ref.~\onlinecite{N2vibCX}, and account for electron impact excitation and de-excitation of nitrogen vibrational levels. Since the vibrational excited states can make an appreciable fraction of the N\textsubscript{2} density, as an approximation, reaction 1 from Table~\ref{tab:tableN2} is used to describe elastic scattering off of the $\nu=1$ and $\nu=2$ excited states in reactions F7 and F8.

\begin{figure}[hbt!]
\centering
\includegraphics[width=.5\textwidth]{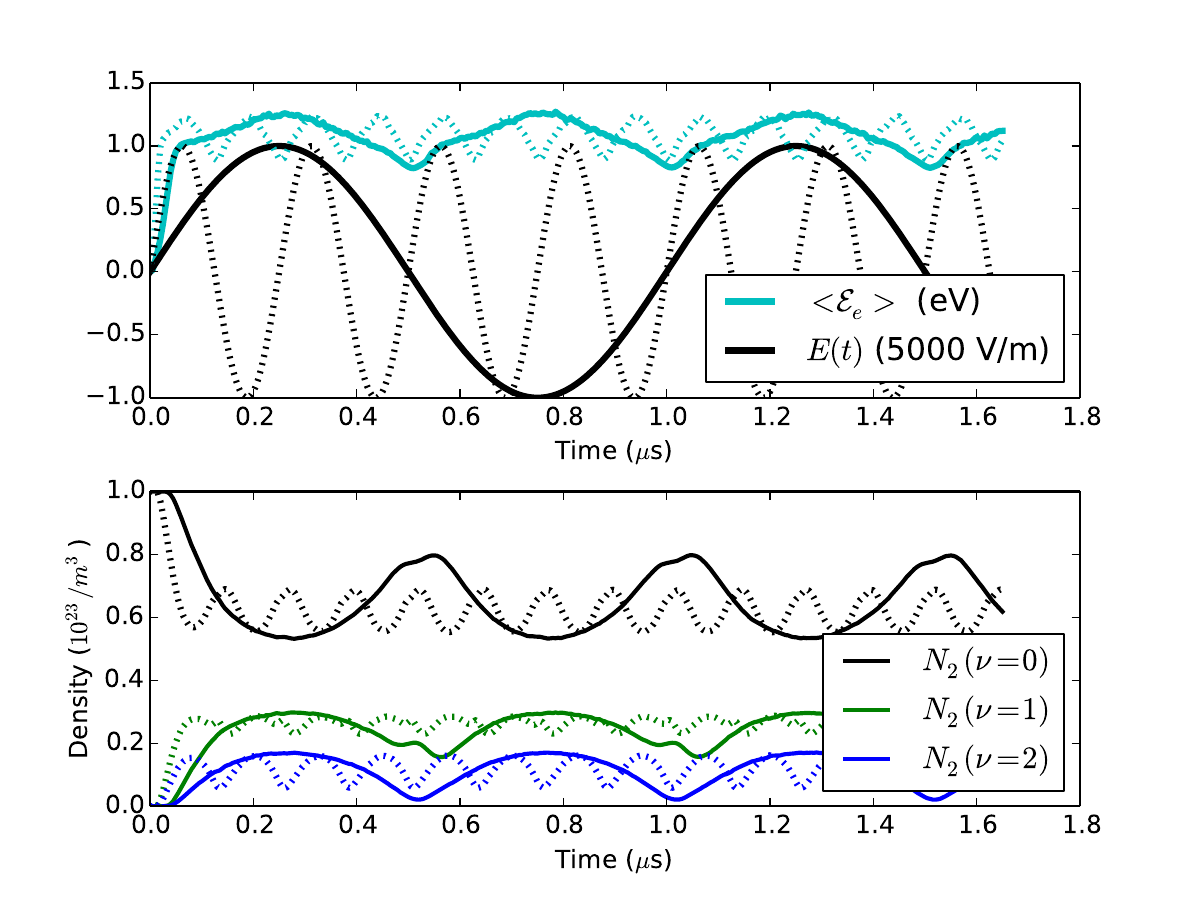}
\caption{\label{N2vibsim}Top: The time dependent driving electric field and resulting mean electron energy in the vibrational N\textsubscript{2} model for electric field drive frequencies of 1~MHz (solid line) and 4~MHz (dashed line). Bottom: The corresponding time-dependent densities of the $\nu=0$--2 vibrational excited states of N\textsubscript{2}.}
\end{figure}

The simulations are carried out with an initial N$_2(\nu~=~0)$ density of $10^{23}~\textrm{m}^{-3}$ in a $1~\mu\textrm{m}^3$ box (100000~particles) and an electron density of $6\times10^{21}~\textrm{m}^{-3}$ (6000~particles). The system is driven with a sinusoidal electric field with a magnitude of $E= 5000$~V/m and a frequency of either 1 or 4~MHz. The electric field and mean electron energy for both the 1 and 4~MHz cases is shown in the top panel of Fig.~\ref{N2vibsim}. As the electric field varies so does the mean electron energy, changing the relative rates of excitation and de-excitation reactions (determined by the cross section and electron VDF) and hence the equilibrium population density for that mean energy. As the field oscillates, the population densities contain some memory of prior configurations that depends on prior values of the electron VDF. These simulations demonstrate the suitability of the DSMC method for problems that track reactants and products and their velocity-dependent interactions.

\begin{table}
\caption{\label{tab:table2} N\textsubscript{2} Vibrational Excitation Model.
%\hl{MZ: Finish excitation thresholds}
}
\centering
\begin{tabular}{lccc}
\hline
Number & Reaction & $\Delta E$ (eV) & Ref.[]\\
\hline
F1&e + N$_2$ $\to$ e + N$_2(\nu=1)$&0.275&\cite{N2vibCX,FLINDERS}\\ 
F2&e + N$_2$ $\to$ e + N$_2(\nu=2)$&0.59&\cite{N2vibCX,FLINDERS}\\
F3&e + N$_2(\nu=1)$ $\to$ e + N$_2$&-&\cite{N2vibCX,FLINDERS}\\
F4&e + N$_2(\nu=2)$ $\to$ e + N$_2$&-&\cite{N2vibCX,FLINDERS}\\
F5&e + N$_2(\nu=1)$ $\to$ e + N$_2(\nu=2)$&0.315&\cite{N2vibCX,FLINDERS}\\
F6&e + N$_2(\nu=2)$ $\to$ e + N$_2(\nu=1)$& -&\cite{N2vibCX,FLINDERS}\\
F7&e + N$_2(\nu=1)$ $\to$ e + N$_2(\nu=1)$&-&\cite{TRINITI}\\
F8&e + N$_2(\nu=2)$ $\to$ e + N$_2(\nu=2)$& -&\cite{TRINITI}\\
\hline
\end{tabular}
\end{table}

%\begin{figure}[hbt!]
%\centering
%\includegraphics[width=.5\textwidth]{VibrationalDensity.pdf}
%\caption{}
%\end{figure}

\subsection{Paschen Curve Calculation\label{sec:PaschenCurve}}
The purpose of this section is to demonstrate the calculation of the breakdown voltage of N\textsubscript{2} gas using slight modifications to the code,
which allow the calculation of
 breakdown voltage characteristics 
 that reproduce textbook Paschen curve results\cite{https://doi.org/10.1002/andp.18892730505,lieberman2005principles}. While the modifications to the code are not part of the main branch of the released version, the level of accuracy achieved suggest that these methods provide one possible solution of how to model boundary loss and source terms in a 0D particle simulation.   

The Paschen curve relation for the breakdown voltage of gases has been well known for more than a century\cite{https://doi.org/10.1002/andp.18892730505}. Briefly, the breakdown voltage for the formation of a self-sustaining discharge between a cathode at location $z=0$ and an anode at location $z=d$ occurs if sufficient secondary electron emission (SEE), usually resulting from ion impact at the cathode, can be obtained to sustain further ionization of the gas. Assuming a constant ionization coefficient in the gap, the electron flux at position $z$ is $\Gamma_e(z)=\Gamma_e(0)\exp(\alpha d)$. In the steady-state condition where the discharge is sustained, the electron and ion charge leaving the system are equal. With this consideration, the ion flux at the cathode is given by $\Gamma_i(0)=\Gamma_e(0)\{\exp(\alpha d)-1\}$.
The secondary electron flux at the cathode needed to sustain the discharge is $\Gamma_e(0)=\gamma_{\rm SEE}\Gamma_i(0)$, where $\gamma_{\rm SEE}$ is the SEE coefficient. This leads to the well-known threshold condition for breakdown of the gas: $1/\gamma_{\rm SEE}+1=\exp(\alpha d)$. This condition depends on the gap distance, the $E/N$ dependent value of the ionization coefficient, and the surface SEE. The final step needed to obtain the well known result is to note that the Townsend ionization coefficient has the form $\alpha/p=A\exp(-Bp/E)$ over a wide range of $E/N$, 
%\hl{MZ: Is the rest of this sentence correct?}\hlcy{BSS: Yes, this is correct.}
where $p$ is the pressure and $A$ and $B$ are constants. Inserting this into the breakdown condition and using $E=V_{\rm break}/d$ leads to the common form of the Paschen curve for breakdown voltage
\begin{equation}
V_{\rm break}=\frac{Bpd}{\ln(Apd)-\ln(\ln(1+1/\gamma_{\rm SEE}))}.
\end{equation}

The physical mechanism that contributes to the growth of density with the propagation of electrons in a fixed field was considered in the calculation of the Townsend ionization coefficient in Sec.~\ref{sec:HeSwarm}. What remains to model the gas breakdown is the tracking of ions, in addition to the electrons, and the modeling of charge loss and SEE at the $z=0$ and $z=d$ boundaries. 
To do this, we make use of the displacement variable (see Sec.~\ref{sec:use}) that was integrated using the particle velocity at each time step. 
Although this spatial variable does not play a role in the collision dynamics or field update, it plays a critical role in the current collection and determining the rate of removal of charge from the system at anode or cathode, and the rate of generation of secondary electrons. 
The boundary interactions were calculated and particles were removed if the criteria  $z>d$ or $z<0$ were met for a given particle. Furthermore, if the particle had a positive charge, a uniform random number $R$ between 0 and 1 was selected, and if $R<\gamma_{\rm SEE}$, an electron was generated at the surface  with a random direction and energy of 1~eV.
In this role, the spatial variable serves as a way to limit the particle lifetime in the domain, but does not provide additional spatial resolution of the particle densities.

The method above was implemented with the addition of ~50 additional lines of code and was used to calculate the breakdown voltage of nitrogen for anode-cathode gaps of $d = 1$, 4, 8, 16, 32, and 64~mm at a pressure of 15.53~Torr. The simulations have a specified electric field which with the gap length $d$ specifies the anode cathode voltage, $V=E\cdot d$. The SEE coefficient was set to $\gamma_{\rm SEE}=0.1$, a value that is similar to that of many materials\cite{Daksha_2019}.
The simulation is initialized with 100 initial electrons at location z=0 and 500000 initial neutrals in a box of volume  $1~\mu\textrm{m}^3$.

\begin{figure}[hbt!]
\centering
\includegraphics[width=.45\textwidth]{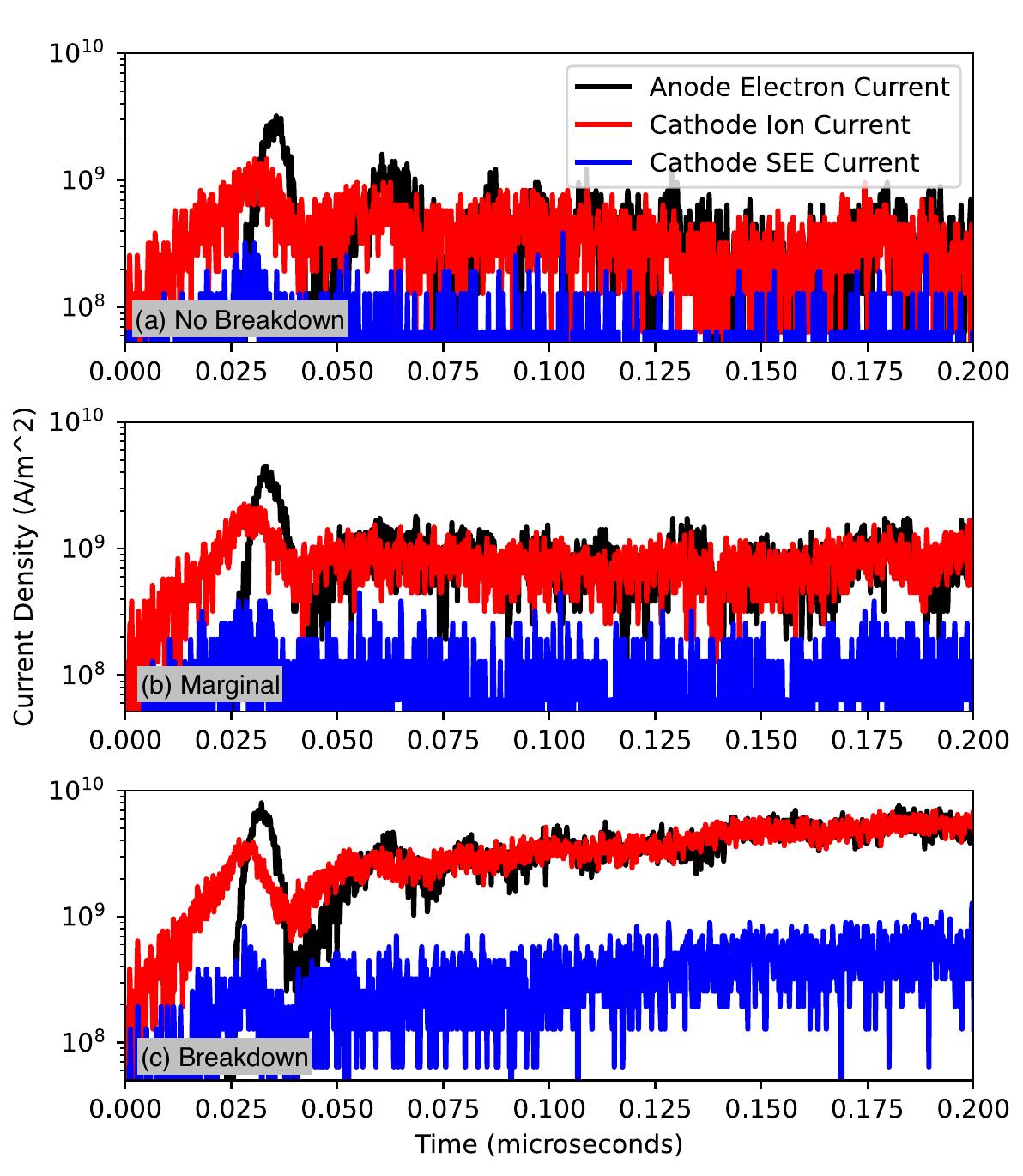}
\caption{\label{fig:currents}The anode electron current, cathode ion current, and cathode secondary electron current densities as a function of time for cases with (a) no breakdown, (b) a marginal sustained discharge, and (c) breakdown of N\textsubscript{2} gas.}
\end{figure}

Figure~\ref{fig:currents} shows three characteristic sets of behavior for the anode and cathode currents that can occur depending on whether or not the breakdown voltage has been achieved. Initially, ion current is collected at the cathode producing secondary electrons at the surface. As the initial electrons increase in displacement they are eventually lost to the anode after some transit time (\tildetext25--40~ns for the case of Fig.~\ref{fig:currents}). If voltage is less than the breakdown voltage the anode electron current peaks when the initial electron avalanche reaches the anode, but by this point there has been insufficient ion production in the gap to sustain further growth of the electron density by SEE at the cathode and the density decreases on average as time increases. The absence of breakdown is clearly identified by the decrease in cathode secondary emission current with time. As shown in Fig.~\ref{fig:currents}(b), very close to the breakdown voltage the electron current at the anode appears to maintain a steady current flow and only increases or decreases on a timescale of microseconds. This case would likely lead to an increase or decrease in anode current with sufficient run time, but the outcome may be sensitive to the stochasticity of the simulation. Slight incrementation of the voltage from case (b) leads to breakdown of the gas and an increasing anode current with time as shown in Fig.~\ref{fig:currents}(c). One hallmark of this behavior is the clear increase in cathode secondary emission current with increasing time. This behavior has been observed and discussed previously in more computationally expensive 1D PIC simulations of the breakdown process\cite{10.1063/5.0051095,Fierro_2017,10.1063/1.4769601}. Here similar behavior is achieved within \tildetext2~hours for the 6 different values of $pd$ using a single processor core.

\begin{figure}[hbt!]
\centering
\includegraphics[width=.5\textwidth]{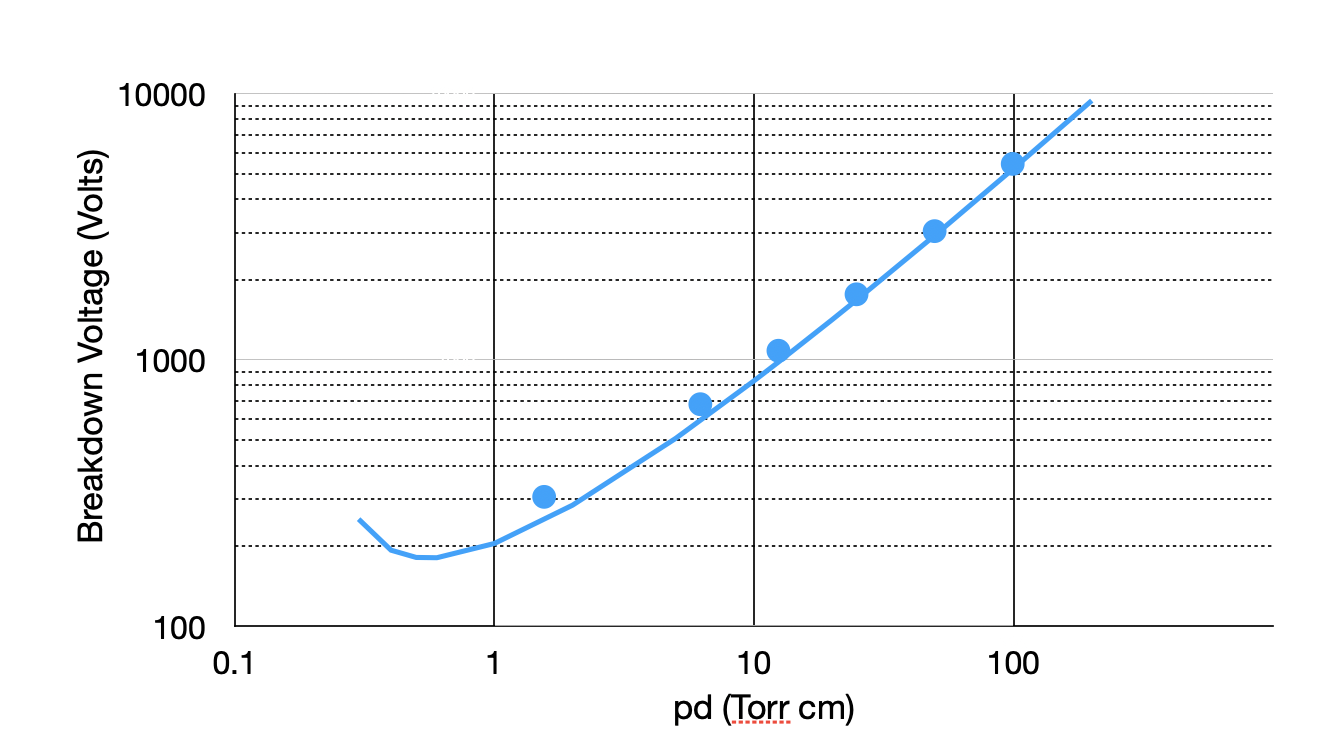}
\caption{\label{fig:paschen} The breakdown voltage of N\textsubscript{2}  determined from simulation compared to the experimentally known Paschen curve values.
}
\end{figure}

Estimation of the breakdown voltage for a given gap distance was achieved by running successive simulations, incrementing the electric field by 5000~V/m until breakdown was observed. Therefore, the precision of the breakdown voltage was $d\times5000$~V/m. For the gap distances involved, individual runs took between 2--15~minutes on a single processor core. Figure~\ref{fig:paschen} compares the results of the simulation with the known values of the Paschen curve for N\textsubscript{2} using the values of $A=11.8/\textrm{cm-torr}$ and $B=325~\textrm{V}/\textrm{cm-torr}$ from Ref.~\onlinecite{lieberman2005principles}. Excellent agreement is obtained above 10~Torr-cm. Disagreement in the lower $pd$ branch is somewhat expected since it is well known that this region requires special attention to the ion-neutral and fast neutral collision models that were not considered in these simulations\cite{10.1063/1.5000387}.

\section{Discussion}

The previous two sections demonstrated several use cases for 0D DSMC simulations, some of which parallel the typical use of multi-term Boltzmann solvers (Sec.~\ref{sec:HeSwarm}), ODE based chemical kinetics solvers (Sec.~\ref{sec:onsager}), and the coupling of the two (Sec.~\ref{sec:kinetics}). Additional example cases demonstrated the coupling of multiple charged, neutral, and gas populations out of equilibrium (Sec.~\ref{sec:gasHeating}), and the use of particle displacement information in the 0D model gain and loss terms (Sec.~\ref{sec:PaschenCurve}). The benefit of the DSMC method for 0D modeling lies in the ability to model all populations with a full kinetic description. 
 These benefits are accompanied by new approaches and challenges for using particles in 0D modeling. 
These can be grouped into two categories described in this section: (1)~{\emph{modeling of populations out of equilibrium}} and (2)~the {\emph{use of displacement information}}. For both of these, current issues, modeling approaches, and possible areas for future development and improvement are discussed.

\subsection{Modeling of Populations out of Equilibrium}
One of the primary advantages of the present method is the ability to couple electron, ion, and neutral populations solely on the basis of the collision interactions and cross sections provided. Section~\ref{sec:gasHeating} already provided one example with the coupling of the electric field, ion, and neutral populations. Another possibility is the coupling of electron and ion behavior due to Coulomb collisions. These could be included in simulations through, for example, the Rutherford cross section, however due to the large number of glancing small angle collisions a model for cumulative small angle collisions\cite{PhysRevE.55.4642} may be more appropriate. Such considerations may be important for the calculation of electron or ion transport coefficients in systems with larger ionization fraction or small values of the plasma parameter $\ln(\Lambda)$ that lead to coupling of the electron and ion components.    

Another possibility for future development involves the coupling of photo-emission and photo-absorption models. As has been demonstrated previously in PIC-DSMC simulations\cite{Fierro_2017}, the emission and absorption of photons by the neutral gas component can be modeled with first-principle-like assumptions that rely on the single atom emission and absorption models. All other properties including the population of states and the Doppler effect of photons emitted by a moving particle are determined directly from the collisions and interactions at the discrete particle level. When added to 0D simulations, these models may be used to predict spectra. 
%from electrons in a given background field. 

The detailed collision modeling of this approach allows one to capture the effect of collision processes on the neutral EDFs that would not be possible with rate-equation-based ODE solvers. 
Situations where these benefits may be of interest are those where the relative energy of the colliding particle pair is important for determining the reaction probability. Examples include the energy dependence of three-body recombination events where slow particles are more likely to combine than fast ones (c.f. Chapter 9.2 of Ref. \onlinecite{lieberman2005principles} and Ref. \onlinecite{10.1063/1.4986529}), the energy dependence of charge-exchange collision\cite{10.1063/1.555889}, and that of vibrational-translational (VT) and rotational-translational\cite{capitelli2013plasma} energy exchange that depletes or populates a particular energy range of the reacting particle's distribution. Models for probabilities of these reaction processes that can be readily cast into the form of velocity dependent cross sections can be used in the present release of ThunderBoltz, while other probability or rate-based models will require the implementation of new collision models. The use of such models is common in DSMC simulations where several particle populations have different temperatures\cite{bird1994molecular},
however the modeling of coupled non-equilibrium populations is generally absent in zero-dimensional plasma modeling where typically only the electrons are assumed to be out of equilibrium with the gas.     

%However, despite the challenges, there are clear benefits to the use of 0D DSMC simulations over rate equation solvers. Use of the DSMC method allows one to account for anisotropy in scattering angle with respect to an applied electric field, a benefit shared with Monte-Carlo Boltzmann solvers\cite{https://doi.org/10.1029/2019JD031564}, the inclusion of charged particle scattering models that couple electron and ion velocity distribution functions\cite{PhysRevE.55.4642}, or the use of photon transport models\cite{Fierro_2017} to create a full 0D collisional radiative model.

%\subsection{Collision Specification via Cross Section Data %\label{sec:detailedCollision}}

%\hlcy{----}

%\hl{MZ: I am not sure this really needs its own section, it could either piggy back on the previous section or maybe some more arguments can be made with collision models. Such as particle flux calculations from fundamental collision models?}

 The primary limitation with the practical implementation of modeling the interaction of non-equilibrium populations is the availability of velocity dependent cross section data for ion-neutral or neutral-neutral reactive collisions. For these interactions data is most commonly available in rate constant form\cite{capitelli2013plasma}. Some solutions for the production of cross section data exist in certain cases.
For example, the generation of cross section data can be done within the framework of Bird's Total Collision Energy (TCE) model\cite{bird1994molecular,doi:10.1063/1.5097706}. The TCE model provides reaction probabilities for temperature dependent Arrhenius rate constants of the form $k_p(T)=aT^b\exp(-E_a/k_bT)$ as $\sigma_{p}(E_c)=\sigma_{\rm VHS}(E_c)P_p(E_c)$ where $P_p(E_c)$ is the reaction probability, $E_c$ is the energy of the colliding particle pair, and $\sigma_{\rm VHS}$ is the variable hard sphere cross section, see Ref.~\onlinecite{doi:10.1063/1.5097706} Eq. (16) and Ref.~\onlinecite{bird1994molecular}. Cross sections of this type can readily be computed and supplied as input to ThunderBoltz. However, these cross sections typically ignore detailed energy dependence of the interaction by presupposing a particular functional form. 

For certain cases, models other than TCE are available. For instance, V-T transfer models for a collision energy dependent probabilities are available\cite{doi:10.2514/2.1181}. In these cases it is more convenient to directly evaluate the collision probability instead of extracting it from the cross section since the needed data is more readily available in probability form. Such functionality is not present in the current version, but is straightforward to implement. Alternatively, in select cases such as for H\textsubscript{2}, vibrational-vibrational (VV) and VT cross section data is available \cite{10.1063/1.4793472}.

\subsection{Use of Displacement Information\label{sec:disp}}
% For ODE rate equation based models, plasma particle generation results from reactions, $p$, with the associated rate constants, $k_p$ that produce species $C$ by collision of particles of type $A$ and $B$. In addition, loss or gain to or from boundaries amounts to the inclusion of terms such as the last two on the right hand side of the particle continuity  equation
% \cite{lieberman2005principles}:
% \hl{MZ: Brett - what is alpha? Does not look like it is needed here}\hlcy{BSS: It was a stoichiometric coefficient, but it could just be combined into k here. }

 For ODE rate equation based models, plasma particle generation results from reactions, and loss or gain to or from boundaries amounts to the inclusion of terms such as the last two on the right hand side of the particle continuity  equation
  \cite{lieberman2005principles}:
\begin{equation}
\frac{dn_C}{dt} = \sum_p  k_p n_{A} n_{B}-(\textrm{Flux out})_C\frac{\mathcal{A}_{\rm out}}{V}+(\textrm{Flux in})_C\frac{\mathcal{A}_{\rm in}}{V},
\end{equation}
where reactions, $p$, with the associated rate constants, $k_p$ produce species $C$ by collision of particles of type $A$ and $B$. 
A mean energy loss per particle can be tracked by an energy equation, and the fluxes can be modified by the use of diffusion constants. Other than these considerations, the particle density is lost or gained with no consideration for the details of the particular particle energy, time, or location in the system because these quantities cannot be tracked by such descriptions. 

On the other hand, DSMC-based 0D models can make use of a tracked particle displacement variable to construct surface loss or source criteria similar to that used in the Paschen curve calculation in Sec.~\ref{sec:PaschenCurve}. Similar criteria may be used to add or remove gas species at different boundaries and mock up the effect of gas residence time. Likewise, these displacement-based particle loss conditions can be combined with energy criteria to model more sophisticated interactions such as an energy dependent secondary electron or sputter yield. These models preferentially remove fast particles first, since these particles accumulate displacement faster than slow particles.  
\

\section{Conclusion}

In this paper, 0D DSMC simulations were carried out with the new code ThunderBoltz to demonstrate the adequacy of using the method as an alternative to multi-term or Monte-Carlo Boltzmann solvers. We also evaluate its suitability for 0D plasma modeling in place of ODE rate equation based plasma chemical kinetics descriptions. Since all species are treated as particles, the simulation removes the need for an external call to a Boltzmann solver that is needed in the ODE description, and allows all species to be modeled natively. The code was tested against a benchmark test problem with varying values of $E\times B$, transport and rate constants in He gas calculated from Bolsig+ and experimental measurements, and known analytic detailed balance relations for neutral species reactions. Example use cases for 0D modeling were presented and possible extensions and applications were discussed. These demonstrate that the present method is superior to ODE-based 0D models when multiple populations are out of equilibrium, or when particle displacement or energy information is needed to implement boundary models in 0D.

Many of the examples above are included with the released version of the code. For ease of use these have been implemented in python scripts making use of the python-based wrapper. This wrapper contains utilities that generate input files using LXCat formatted data, making the code  accessible to new users. ThunderBoltz is currently serving as a platform to test new PIC-MCC/DSMC collision models that will be presented in other upcoming publications. These new models and other additional features, some of which have been mentioned in this manuscript, will be added in the near future.

\section*{Acknowledgments}
All authors contributed equally to this work. Brett Scheiner thanks Matthew Hopkins for numerous conversations and mentorship on these topics over the past 9 years. The authors also thank Christopher Moore for useful discussions. Ryan Park and Mark Zammit were supported by the U.S. Department of Energy through the Los Alamos National Laboratory
ASC PEM Atomic Physics Project. Los Alamos National Laboratory is operated by Triad National Security, LLC, for the National Nuclear Security Administration of U.S. Department of Energy (Contract No. 89233218NCA000001).

\bibliography{main}

\end{document}